\documentclass [12pt,twoside]{article}           % version LATEX2E
 \usepackage[left,running]{lineno} %running-> toutes les lignes 
\usepackage{setspace}
\countdef\arxiv=201
\ifnum\arxiv=0 \input cpreb.tex \fi

\ifnum\arxiv=1

\usepackage{epsfig,times,lscape}
\usepackage[usenames]{color}

    \ifnum\count102=1

    \fi

    \ifnum\count102=2
\fi

\newcommand{\nc}{\newcommand}

     \ifnum\count101=1
\nc{\qI}[1]{\section{{#1}}}
\nc{\qA}[1]{\subsection{{#1}}}
\nc{\qun}[1]{\subsubsection{{#1}}}
\nc{\qa}[1]{\paragraph{{#1}}}

\def\qpar{\vskip 2mm plus 0.2mm minus 0.2mm}
\def\qL{\hfill \break}
     \fi 

      \ifnum\count101=2
 \nc{\qI}[1]{\parindent=0mm \vskip 8mm 
{\centerline{\LARGE \color{red}#1}}\vskip 3mm}
\nc{\qA}[1]{\vskip 2.5mm \noindent 
{{\bf\large\color{blue}  #1}} \vskip 1mm \parindent=0mm}
 \nc{\qun}[1]{\vskip 1mm \noindent {\sl \color{blue} #1 }\quad }

\def\qL{\hfill \break}
\def\qpar{\vskip 2mm plus 0.2mm minus 0.2mm}

      \fi
\def\qth{\vrule height 12pt depth 0pt width 0pt}
\def\qtb{\vrule height 0pt depth 5pt width 0pt}

\nc{\qfoot}[1]{\footnote{{#1}}}

      \ifnum\count101=1
\def\qbu{\hfill \par \hskip 6mm $ \bullet $ \hskip 2mm}
\def\qee#1{\hfill \par \hskip 6mm (#1) \hskip 2 mm}
      \fi
      \ifnum\count101=2
\def\qbu{\hfill \par \hskip 4mm $ \bullet $ \hskip 2mm}
\def\qee#1{\hfill \par \hskip 4mm (#1) \hskip 2 mm}
      \fi

\def\qparr{ \vskip 1.0mm plus 0.2mm minus 0.2mm \hangindent=10mm
\hangafter=1}

     \ifnum\count101=1 
 \def\qdec#1{\parindent=0mm\par {\leftskip=2cm {#1} \par}}
     \fi
     \ifnum\count101=2
  \def\qdec#1{\parindent=0mm \par {\leftskip=1cm {#1} \par}}
  
  \def\qcitb#1{\noindent \hbox to 102mm{\hfill \small #1} \vskip 1mm}
      \fi

 \def\qpages#1{\count102=0{\loop\advance\count102 by 1
 \null \vfill\eject \ifnum\count102<#1 \repeat}}

\def\qth{\vrule height 12pt depth 0pt width 0pt}
\def\qtb{\vrule height 0pt depth 5pt width 0pt}

\def\qv{\vskip 0.1mm plus 0.05mm minus 0.05mm}
\def\qhu{\hskip 0.6mm}
\def\qhv{\hskip 3mm}

\def\qhw{\hskip 1.5mm}
\def\qleg#1#2#3{\noindent {\bf \small #1\qhw}{\small #2\qhw}{\it \small #3}\qv }
\newcommand{\promille}{%
  \relax\ifmmode\promillezeichen
        \else\leavevmode\(\mathsurround=0pt\promillezeichen\)\fi}
\newcommand{\promillezeichen}{%
  \kern-.05em%
  \raise.5ex\hbox{\the\scriptfont0 0}%
  \kern-.15em/\kern-.15em%
  \lower.25ex\hbox{\the\scriptfont0 00}}

\fi
\begin{document}
\thispagestyle{empty}

% --------------------------------------------------------------------

      % Hauts de pages et numerotation

          % Remarque: sans le \protect --> message d'erreur (ordre fragile)
\markboth{{\sl \hfill  \hfill \protect\phantom{3}}}
        {{\protect\phantom{3}\sl \hfill  \hfill}}

% -------------------------------------------------------------------
\color{yellow} 
%\hrule height 20mm depth 20mm width 170mm 
\hrule height 10mm depth 10mm width 170mm 
\color{black}

 \vskip -15mm   % pour un titre avec 1 seule ligne
%\vskip -17mm   % pour un titre avec seconde ligne

%\centerline{\bf \Large Cross-species investigation of infant mortality.}
%\vskip 3mm \centerline{\bf \Large Part 1: Motivations and background}
%
% RECHERCHE GOOGLE: ``Congenital anomalies''     -> 2.4 millions
%                   ``Congenital abnormalities'' -> 1.2 millions
%

\centerline{\bf \Large Congenital anomalies from a physics perspective.}
\vskip 3mm 
\centerline{\bf \Large The key role of ``manufacturing'' volatility}
\vskip 5mm
\centerline{\bf \Large }
\vskip 10mm

\centerline{\normalsize
Alex Bois$ ^1 $,
Eduardo M. Garcia-Roger$ ^2 $,
Elim Hong$ ^3 $,
Stefan Hutzler$ ^4 $,
Ali Irannezhad$ ^5 $},
\qL
\centerline{\normalsize
Abdelkrim Mannioui$ ^6 $,
Peter Richmond$ ^7 $,
Bertrand M. Roehner$ ^8 $,
St\'ephane Tronche$ ^9 $
}

\vskip 5mm
\large

%{\bf \color{red} SUMMARY}\qL
%{\bf \color{blue} Background:} \quad
%{\bf \color{blue} Aim:} \quad 
%{\bf \color{blue} Method:} \quad 
%{\bf \color{blue} Findings:}\quad 
%{\bf \color{blue} Conclusion:}\quad 
%
                              
\vskip 5mm
\centerline{\it \small Version of 3 May 2019}
%\centerline{\it \small Provisional. Comments are welcome.}
\vskip 3mm

{\small Key-words: Congenital anomalies, malformations,
infant mortality, manufacturing defects} 

\vskip 3mm

{\normalsize
1: Aquatic facility, Pierre and Marie Curie Campus, Sorbonne University,
Paris, France.\qL
Email: alex.bois@upmc.fr\qL
2: Institut Cavanilles de Biodiversitat I Biologia Evolutiva,
University of Val\`encia, Spain.\qL
Email: eduardo.garcia@uv.es\qL
3: Neuroscience Laboratory, Sorbonne University and INSERM
(National Institute for Health and Medical Research).\qL
Email: elim.hong@inserm.fr\qL
4: School of Physics, Trinity College, Dublin, Ireland.\qL
Email: stefan.hutzler@tcd.ie\qL
5: School of Physics, Trinity College, Dublin, Ireland.\qL
Email: irannezhad.a@gmail.com\qL
6: Aquatic facility, Pierre and Marie Curie Campus, Sorbonne University,
Paris, France. \qL
Email: abdelkrim.mannioui@upmc.fr \qL
7: School of Physics, Trinity College Dublin, Ireland.\qL
Email: peter\_richmond@ymail.com \qL
8: Institute for Theoretical and High Energy Physics (LPTHE),
Pierre and Marie Curie campus, Sorbonne University,
Centre de la Recherche Scientifique (CNRS).
Paris, France. \qL
Email: roehner@lpthe.jussieu.fr\qL
9: Aquatic facility, Pierre and Marie Curie Campus, Sorbonne University,
Paris, France. \qL
Email: stephane.tronche@upmc.fr
}
\vfill\eject
%---------------------------------------

\large

{\bf Abstract}\qL
Genetic and environmental factors are traditionaly seen as the 
sole causes of congenital anomalies. In this paper we
introduce a third possible cause, namely
random ``manufacturing'' discrepancies with respect to 
``design'' values. A clear way to demonstrate the existence
of this component is to ``shut'' the two others and to see
whether or not there is remaining variability.
Perfect clones raised under
well controlled laboratory conditions fulfill the conditions
for such a test.
Carried out for four different species,
the test reveals a variability remainder of the order
of 10\%-20\% in terms of coefficient of variation.
As an example, the CV of 
the volume of E.coli bacteria immediately after
binary fission is of the order of 10\%.\qL
In short,
``manufacturing'' discrepancies occur
randomly, even when no harmful mutation
or environmental factors are involved.
If the pathway is particularly
long or requires exceptional accuracy, 
output dispersion will be high and
may lead to malformations. 
This effect will
be referred to as the {\it dispersion effect}.
We conjecture that it will be particularly significant when
major changes occur; this includes the early phase
of embryogenesis or the steps
leading from stem cells to differentiated
(organ-specific) cells.
\qL
The dispersion effect not only causes malformations
but also innocuous variability. For instance 
monozygotic (MZ) twins resemble each other 
but are not strictly
identical. It is not uncommon to see only one 
of the twins of a MZ pair showing 
a congenital defect (see Appendix A).
\qL
Not surprisingly, there is a strong connection
between congenital defects and
infant mortality. In the wake of birth 
there is a gradual elimination of defective units 
and this screening accounts for
the post-natal fall of infant mortality.
For reasons which are not yet fully understood,
this fall continues until the age of 10 years. 
Neither do we understand why, as a function of age,
the downward trend of human infant mortality 
follows a power law
with an exponent around 1 (whereas for fish 
it is about 3, see Bois et al. 2019a).
Apart from this trend,
post-natal death rates also have humps and peaks associated 
with various inabilities and defects.\qL
In short, infant mortality rates 
convert the case-by-case and mostly qualitative problem of congenital
malformations into a global quantitative effect which, so to say,
summarizes and registers what goes wrong in the embryonic phase.
\qL
Based on the
natural assumption that for simple organisms (e.g. rotifers)
the manufacturing processes are shorter than for
more complex organisms (e.g. mammals), fewer
congenital anomalies are expected. 
Somehow, this feature should be visible on the infant mortality
rate. How 
this conjecture can be tested
is outlined in our conclusion.

\vfill\eject
%----------------------------

\count101=1  \ifnum\count101=1

{\it \Large \color{blue} Contents}
\qpar

\qun{Introduction: the ``manufacturing dispersion'' effect}\qL
{\small \color{black} 
Variability in biochemical reactions \qL
From technical systems to living organisms: a physics perspective\qL
Broad reach of congenital anomalies \qL
Randomness of the dispersion effect \qL
Rationale for an output dispersion effect \qL
Control procedures \qL
Why defect statistics give a biased picture \qL
Are complex organs more affected by output dispersion? \qL
The most frequent defects have close links with normality \qL
Weak role of genetic factors in birth defects \qL
Identification of output dispersion through phenotype variations \qL
Outline of the paper
}
\qun{Fault-tolerant design}\qL
{\small \color{black} 
Tolerance issues in the industrial production of mechanical devices
\qL
The tolerance system as a way to mitigate the effects of
manufacturing defects \qL
Output dispersion in biological systems
}
\qun{Salient features of embryonic mortality}\qL
{\small \color{black}
Implication of geometrical abnormalities for development
of the embryo \qL
Age-dependent embryonic mortality \qL
General observations  about embryonic mortality \qL
Avian species. Fish species. Human fetal deaths \qL
How can one explain that the death rate is highest
at the beginning of embryogenesis? \qL
A conjecture about embryogenesis in unicellular organisms \qL
Influence of temperature on hatching rate
}
\qun{Salient features of the two phases of human mortality}\qL
{\small \color{black}
Human infant mortality for all causes of death \qL
The age of 10 seen as an equilibrium point between screening and
  wear-out \qL
Infant mortality for specific causes of death
}
\qun{Conclusion}\qL
{\small \color{black}
Main results \qL
Rationale for cross species comparisons
}
\qun{Appendix A. Estimating the strength of genetic factors}\qL
{\small \color{black}
Mutations and repair mechanisms \qL
The twin methodology for assessing the strength of genetic
factors \qL
Strength of genetic factors in cancer
}

\vfill\eject

\fi

%----------------------
\large
 
\qdec{\it This paper is the first leg of an exploration
in three parts; the two others are Bois et al. (2019a,b).
Despite the connections the three papers can be read
independently from each other.}
\vskip 4mm

\qI{Introduction: the ``manufacturing dispersion'' effect}

In a characteristic way
the abstract of a recent paper about birth defects 
begins with the following sentence:
``The causes of birth defects are complex and include genetic and
environmental factors and/or their interactions'' (Chen et al. 2018).
In other words, genetic and environmental factors are seen as
the only sources of birth defects. Here we add 
a third source referred to as a ``manufacturing dispersion'' effect.
Its introduction is motivated by several reasons
which are outlined in coming subsections. 

\qA{Variability in biochemical reactions}

Thousands of biochemical reactions are required for the growth
of any living organism even if it is a single cell.
Taken together they constitute what one
may call a manufacturing process. For each of these reactions
there is a set of optimal parameters in terms
of temperature, pH, concentration
of enzyme-catalyst, orientation and shape of interacting
molecules and so on.
It is clear that mutations and environmental factors
may disrupt this process. However, in the present paper
we develop the idea that even if all parameters are set at their
optimal design values%
\qfoot{In practice this means: (i) a time interval
sufficiently short to ensure that the likelihood of
mutations is negligible compared to other reactions.
(ii) constant optimal environmental
conditions of the kind maintained in controled laboratory
experiments.}
nevertheless
there will be a dispersion of the outcomes. It
has four main causes. (i) Initial conditions
may not be identical. (ii) Even if initial
conditions are very similar, there will be ``butterfly effects''
(due to the nonlinearity of the reactions) which will greatly
amplify any initial dissimilarities no matter how tiny.
(iii) The parameters defining the reactions are never
{\it exactly} at their optimum values.
(iv) Random quantum fluctuations cannot be avoided.
Note that
this last effect is probably smaller than the others.
\qL
Even if at each step the
volatility is small, a succession of steps will result in
a cumulative effect which, eventually, may lead to
noticeable congenital anomalies. 
\qpar

As an illustration of this kind of variability
consider an observation made at the level
of individual cells. According to a recent study 
(Wallden et al. 2016, p.729,733, Fig. 4B,C),
isogenic E. coli cells (i.e. having same genotype)
growing in a uniform and invariable environment 
display significant variability in volume at birth (i.e.
volume immediately after binary fission)
and in individual growth rates. 
The coefficients of variation are 
fairly substantial, of the order
of CV$ \simeq 10\%,\ 20\% $ respectively. 
Actually, variability at cell level has already been
recognized and studied (at least qualitatively) in the 
1910s and 1920s 
as will be documented later on.

\qA{From technical systems to living organisms: a physics perspective}

In this paper we examine biological systems from the
perspective of reliability engineering.
Such a comparative approach 
is rather uncommon in biology; in contrast, comparative
analysis plays a key-role in experimental physics. Therefore,
it is perhaps not 
surprising that it is tried by physicists and biologists
who share a similar turn of mind%
\qfoot{Not long ago, in an email of 31 December 2018,
Prof. Bert Vogelstein,
a biologist renowned for his work on
cancer told us: ``We need
more physicists thinking about cancer''. 
Such a statement was certainly an encouragement.}%
.
\qpar

Why should it be useful to establish a link between
technical and living systems?
In physics it is natural
to take systems that we understand pretty well as starting
points for the investigation of phenomena that remain
mysterious%
\qfoot{Many such cases can be found throughout the
history of physics. One of the most recent examples
is how electromagnetism, more precisely quantum electrodynamics
(QED)
was used as a guide for building a theory of strong
interactions, namely quantum chromodynamics (QCD).}%
.
\qpar

One should not focus only on
similarities, differences may also be revealing.
A rather obvious illustration is that, whereas
in engineering the duplication of critical components
is a common technique for improving reliability,
mammals have only one heart not to
speak of many other vital functions for which there is no
backup. For instance,
urinary retention can occur for many reasons
whether physiological or neurological
and, if not remedied, 
may lead to death within a few hours. Yet, there is no
backup mechanism. We are told that Tycho Brahe, one
of the founding fathers of modern astronomy, died
that way. This example is of interest because, whereas 
adding a second heart
would require a considerable design change,
creating a supplementary bladder outlet would be a fairly
simple matter.

\qA{Broad reach of congenital anomalies}

\qun{Malformations versus deficiencies}
This paper is mainly about congenital anomalies
We prefer this expression to birth defects for
two reasons:
(i) Many anomalies do not appear in the form of
malformations but as deficiencies, e.g. insufficient
production of insulin in Type 1 diabetes. 
(ii) Many congenital anomalies do not appear at birth
nor even in childhood but much later in the course
of life; anomalous heart valves are an example that will
be discussed later on. Behavioral anomalies may also appear
only later in the course of life.
Having said that, we will sometimes
also use ``birth defects'' which has the advantage
of being shorter.

\qun{Anomalies of the immune system}
It should
be observed that in fact it is difficult to separate
mortality due to congenital anomalies
from other causes of death.
Even cancer or
mortality from infectious diseases may be attributed
to congenital anomalies of the immune system. 
In this respect one should remember that even in major 
epidemics such as the Spanish influenza pandemic
of October-November 1918
less than 10\% of the population was affected in the
sense of being hospitalized and only about 0.4\% died
which means that most persons were protected by their
immune system. Only a few were not.

\qun{Behavioral anomalies}
The behavior of living organisms is to a large extent
genetically controled. As an example consider the case of 
a broody hen. From the eggs laid by the hen to the hatching 
of chicks 21 days later there is a succession of steps which
is quite remarkable. 
\qpar

(i) The process starts when the eggs are fertilized
by the rooster inside the hen's body. 
(ii) {\it Physiological changes.} The
beginning of the process is also marked by
physiological changes: the body temperature of the hen 
increases and the feathers under her body fall off%
\qfoot{Injections of the hormones prolactin, luteinizing, and
oestradiol to non-broody hens induces broodiness.}%
.
(iii) {\it Making a nest.} The hen makes a nest about
5cm deep by scratching the ground.
(iv) {\it Storage of the eggs.} 
As the hen will brood a set of about 6 to 10 eggs,
over several days she will lay eggs and store them in the nest.
A delay of up to 6 days will make little difference in hatching 
time. 
As soon as an egg has been laid, it will cool
down and the content will contract whereby the 
air cell is created. It will play a crucial role during
hatching because it is always on this side that the chicks
will pierce the shell.
(v) {\it Sitting on the eggs.} While sitting on the eggs, the
hen will have to turn them in order to prevent the embryonic chicks
from sticking to the shell. 
 As
well as turning them she will also move the eggs on the
outside of the nest into the middle and the middle ones out so all are
evenly warmed. A graph presented in the section on embryogenesis
shows that in terms of temperature there are very strict 
requirements.
(vi) {\it Cleaning the nest.} The hen will have to keep the
nest clean and tidy which, in particular,
means that non-fertilized or broken eggs must be discarded.
(vii) {\it Taking breaks.} The hen will leave the eggs one,
two or three
times a day (each time for about 15 minutes) to find food and
water and to defecate. Nonetheless, a hen will usually lose weight
while brooding.
(viii) {\it Last three days.} 
Toward the end of the incubation and particularly during the
last three days the embryos start to produce significant levels 
of metabolic heat. Therefore, brooding should be relaxed.
When the chicks start to break their shells the hen must give
them enough room. 
(ix) After hatching the chicks remain close to or underneath the
hen thereby sharing her body heat. For the same reason, after hatching
in an incubator chicks are kept warm by infrared lamps.
They are fully feathered only at six weeks of age.
In natural conditions the hen will show her chicks how to identify 
and peck food.
\qpar

It is by purpose that we have described this process
in some detail to show how easily it can be disrupted
or become sub-optimal (in the sense of a reduced hatching rate).
Usually the disruptions which may trigger anomalies
remain hidden to the outside observer. On the contrary,
in the brooding process inappropriate 
environmental conditions which at each step may 
derail the process occur in full view 
and can be identified%
\qfoot{In other words, the analysis of the brooding
process offers
an excellent observational opportunity to explore the interaction
between genotype and environmental conditions.
To our best knowledge, this field of research
has not yet been explored in a systematic way.}%
.
This gives an intuitive view of the notion of manufacturing
volatility. Just as for phenotype
characteristics, there is also a substantial
variability in brooding ability and behavior. 
Some hens are very good at brooding while others are not%
\qfoot{A simple test consists in putting an egg in front of a hen.
If she pulls the egg under her she may be well ``gifted''.}%
.
For instance, first time brooders might not stay 
broody for very long.
\qpar

In the same way as birth defects are nothing but amplified forms
of normal variability, similarly some forms of behavior become
sufficiently extreme to be labelled as ``abnormal''. Here
are a few examples.
\qbu Sometimes, hens will go broody without eggs underneath them.
In some cases they may continue to sit on empty nests 
for 2 or 3 months.
\qbu In many bird species males and females alternate sitting
on the eggs. Curiously, the same behavior was observed 
for two hens (Buibaku et al. 2010).
During the day they were alternating:
one was incubating from morning till noon 
while the other was out to eat, drink, dust bath and
rest. Roles were interchanged around midday.
During the night the two hens were jointly brooding.
However, the result was not satisfactory in the sense that
of the 22 eggs they were brooding only one was able to hatch.

\qA{Randomness of the dispersion effect}

The main defining characteristic
of the dispersion effect is its randomness.
However, this word does not mean that anything can 
happen and that nothing can be predicted. In fact, there are
predictable consequences. 
For instance the dispersion does
not manifest itself in the same way
in a process that requires high accuracy%
\qfoot{For instance, for the eyes even a small 
disymmetry may result in strabismus. For the ears
synchronization requirements are
less critical.}
than in one which does not. Several illustrative examples
are described below.
Whether the dispersion occurs at the beginning or
at the end of a pathway%
\qfoot{In molecular biology the term ``pathway'' has a 
technical meaning in reference with the expression of
genes. Here, we use the word more broadly as referring to
a succession of steps realizing a given
function. It can be a cascade of chemical reactions or also
a succession of actions. An illustration is the feeding function
which requires an organism to see the prey, then to identify 
and catch it and finally to eat and digest it.}
will also make a difference.

\qA{Rationale for an output dispersion effect}

There are several motivations for introducing the dispersion 
effect.
\qee{1} {\bf Most birth defects are unexplained.}\quad
For most birth defects the factor
responsible is not known. A recent publication
in the ``British Medical
Journal'' (Feldcamp et al. 2017) tells us that in a total
of 5,504 birth defects in 270,878 children born in the 
state of Utah in 2005--2009, the etiology is unknown for
3,390 which represents 80\% of the cases. Of the 1,104 cases
for which the etiology is known, 844 are due to
chromosomal abnormalities which are mostly trisomy 13, 18 and 21.
In our conception most defects occur
randomly, so it is hardly surprising that 
many remain unexplained.
\qee{2} {\bf Variability in true twins.}\quad
Many articles (e.g. Ahmed et al. 2017)
give the (misguided) impression that most
malformations can be attributed to specific genes.
If this were true, the twins of monozygotic pairs
would have the same birth anomalies. In fact, as shown
in Appendix A, the discordant cases (where the two twins
do not have the same defect) are 4 times more frequent
than the concordant cases (where they share the same defect). 
\qL
At this point it is necessary to say a word about
epigenetic changes, a notion which refers to how genes
are expressed rather than to their identity.
The present-day consensus is that to be considered
as epigenetic a trait has to be heritable at least
for a number of generations. This is certainly a wise rule for
otherwise any difference occurring between true twins could
(somewhat arbitrarily) be attributed to epigenetic factors. 
\qee{3} {\bf Variability of offspring in uniparental 
reproduction.}\quad 
Inheritance from two parents is a difficult problem.
The study of true twins is one way to overcome this
difficulty. The study of reproduction from a single parent
is another.
Uniparental reproduction was much studied 
between 1900 and 1930 particularly at the ``Zoological
Laboratory'' of John Hopkins University; see the
studies of  
Ruth Stocking (1913,1915), Ralph Middleton
(1915), Herbert Jennings (1916), Bessie Noyes (1922).
Uniparental reproduction (also called asexual reproduction) 
occurs in two cases. \qL
The simplest is the
reproduction by fission of unicellular organisms. 
In her thesis (Noyes 1923) Bessie Noyes cites four 
species of protozoans for which inheritability was
studied. \qL
The same kind of investigation can be made for
multicellular organisms (i.e. metazoans) with 
uniparental reproduction.
For instance, in rotifer species 
during its life time of a few days one female can generate
successively of the order of 10 offspring. Although they
are in a sense clones of their mother, they present a
substantial variability (Noyes 1923).
\qL
It is true that one can never exclude that a somatic mutation
(i.e. a DNA alteration) occurred during 
the embryogenesis of offsprings. Yet, it is
well known that errors in protein synthesis
are far more frequent than errors in DNA replication
(Drummond et al. 2009).
\qee{4} {\bf Dispersion of outputs.}\quad
The three previous points explain that there is room
for a third source of birth defects but it does not
describe what this source could be. It is simply the
fact in any manufacturing process%
\qfoot{To use for living organisms 
the expression ``manufacturing process'' may seem odd.
However, our objective
is precisely to watch living systems from the perspective
of technical reliability science.}
there are two parts:
(i) The design phase (ii) The implementation of the design.
For living organisms it is the DNA-RNA code which represents
the design instructions destined to the
manufacturing process.\qL 
In real life, a design is 
never carried out with absolute accuracy. If a table is designed
with a width of 3m, in reality its width will be 
comprised between $ W_1=2,999 $mm  and $ W_2=3,001 $mm. 
For most practical
usages such small discrepancies are of no
consequence. However, if one wants to bring the table into
a room whose door has a width of 3m, then the $ W_1 $ table 
will get through whereas the $ W_2 $ table will not.\qL
This is a static view. As soon as there is a nonlinear
process evolving in time (which is the case of most biochemical
reactions) there will be butterfly effects through which small
initial differences are amplified. 
\qee{5} {\bf Crucial role of early discrepancies}\quad
In 2015 it was shown that mutations which eventually
lead to cancer cells
may occur at different stages of the transformation of
undifferentiated stem cells into mature
differentiated cells (Tomasetti et al. 2015).
This discovery provided a natural explanation for the
fact, known since the 1920s (Greenough 1925, Patey et al. 1928),
that cancer cells which have a low 
degree of differentiation are also the most malignant,
that is to say, result in early recurrence and death.
Indeed, a mutation occurring early in the differentiation
process will impact and derail all following stages.
\qpar

There is a similar feature with the embryo itself in the sense
that organs in the earliest stage of their development
are most sensitive to teratogenic (i.e. causing  developmental 
malformations) factors at
the time of their appearance. 
This point is shown very
clearly in a paper by Uchida et al. (2018); in this study
various shocks (e.g. heat shocks) were applied in different
stages of the embryo development of zebrafish, frogs and 
chicken. In all cases embryonic lethality was the most
severe when the shock was applied in the earliest stage.
\qpar

This observation has a natural interpretation in the
manufacturing framework; it says
that a small defect in a component $ A $ used
in the early stages of a production chain may have 
quite detrimental 
consequences because it may hinder the appropriate 
working of components introduced later on in the process
and with which $ A $ is functionally related. 
\qpar

The manufacturing conception developed in this paper
is consistent with (yet broader than) the mechanism
identified in Tomasetti et al. (2015) and which the authors
describe as follows: 
\qdec{``The concept underlying the current work 
is that many genomic changes
occur simply by chance during
DNA replication rather than as a result of carcinogenic factors.''
Therefore, one expects a correlation between 
``the lifetime number of divisions 
among the stem cells within each organ and the lifetime 
risk of cancer arising in that organ.''}
\qpar

Each division bringing about a further step in the differentiation
process also represents a new manufacturing challenge which
makes it more prone to output dispersion than mere divisions
into identical daughter cells. Whether the discrepancy occurs by
mutation or by output dispersion, its impact will be more severe
if it occurs early in the differentiation chain.
\qpar

In medical language, such early cell anomalies are labeled
as pre-cancerous conditions. They are characterized
by the presence of abnormal cells, yet in low proportion
and in shapes which are not very different from
normal types.
\qpar

In the manufacturing of living organisms
mechanical operations play a role (see below) but most
pathways consist of a succession of chemical reactions.
The previous argument remains valid however.
Conditions of concentrations, temperature,
acidity or other parameters
are never 100\% optimum; as a result, the outputs
will have a dispersion around optimal design values.
\qpar

%
%%-----------------------------------------------
\begin{table}[htb]

\small
\centerline{\bf Table 1: Incidence of birth defects in 
high accuracy processes.}

\vskip 5mm
\hrule
\vskip 0.7mm
\hrule
\vskip 0.5mm
$$ \matrix{
\hbox{Birth defect}\hfill & \hbox{Description} \hfill & \hbox{Prevalence}\cr
\qtb
\hbox{}\hfill & \hbox{} & \hbox{(per 1,000)}\cr
\noalign{\hrule}
\qth
\hbox{\color{blue} ``All'' birth defects}\hfill & \hbox{} 
& 30\cr
\hbox{}\hfill & \hbox{} & \hbox{}\cr
\hbox{\color{blue} Cases with ``geometrical'' defects}\hfill & \hbox{} 
& \cr
\hbox{\quad Strabismus} \hfill & \hbox{Eyes not properly synchronized}\hfill
& 20 \cr
\hbox{\quad Heart valves defects}\hfill & \hbox{Abnormal joints of
  cuspids} \hfill
& 10 \cr
\hbox{\quad Cleft palate and/or cleft lip}\hfill & \hbox{Facial sheets
  do not join well} \hfill
& 1\cr
\hbox{\quad Spina bifida (open)}\hfill & \hbox{Defect in spine closure} \hfill
& 0.4\cr
\hbox{\quad Spina bifida occulta}\hfill & \hbox{Slight defect in spine
  closure} \hfill
& 150\cr
\hbox{}\hfill & \hbox{} & \hbox{}\cr
\hbox{\color{blue} Among children with trisomy 21}\hfill & \hbox{} 
& \cr
\hbox{\quad Strabismus} \hfill & \hbox{Eyes not properly synchronized}\hfill
& 350 \cr
\qtb
\hbox{\quad Heart} \hfill & \hbox{Serious congenital heart defects}\hfill
& 400 \cr
\noalign{\hrule}
} $$
\vskip 1.5mm
Notes: Prevalence is defined as the
total number of births affected by the problem in a
time interval of several years
compared to the total number of live births in
the same time interval. All these cases are characterized
by ``mechanical'' or ``geometrical'' defects. 
The cuspids designate the leaflets which form the valve.
In most valves there should be three leaflets; when
two leaflets stick together it is a bicuspid defect.
There can also be 1 or 4 cuspids but these defects are
fairly rare. Incidentally,
the fact that the prevalence of the four
causes mentioned is higher than the ``all defect''
prevalence estimate shows that the ``all defect'' notion does
not include some light cases (e.g. light strabismus
or spina bifida occulta)
or defects which manifest themselves only later
in the course of life (e.g. light valve defects). 
Most often spina bifida occulta (i.e. not visible)
causes no symptoms and is only identified through X-ray imaging.
\qL
Trisomy 21 (that is to say three chromosomes number 21 instead of two)
results in over-production of the proteins 
under the control of the 310 genes located on this chromosome.
This disrupts many mechanisms and particularly those
requiring high accuracy: brain (100\% are more or less affected), 
heart (40\% serious congenital anomalies), 
eyes (strabismus affects 35\%), ears (hearing loss affects 70\%).  
\qL
{\it Source: Child health, USA 2014, Table 1: National prevalence
estimates of selected major birth defects; 
Gunton et al. (2015); for spina bifida occulta: estimate of
the ``National Institute of Neurological Disorders and Stroke''.}
\vskip 2mm
\hrule
\vskip 0.7mm
\hrule
\end{table}
%%-----------------------------------------------
%
\qee{6} {\bf Critical processes are the most affected.}\quad
Under the term ``critical processes'' we understand processes
which require high synchronicity and accuracy. Whenever two sheets
must grow at the same speed in order to join seamlessly, even a slight
discrepancy may affect the closure. Examples of
defects of this kind are:
\qbu Spina bifida, a  defect of closure around the spine.
From the open to the closed form 
there is a broad range of severity for this defect.
Spina bifida occulta is a closed form which is quite
frequent; it affects 15\% of newborn according to estimates
but causes no sypmtoms. About this case one can read the following
assessment: ``The exact causes of spina bifida occulta are not well
understood. Both genetic and environmental factors seem to play a
role''. Our thesis is that there are {\it no} causes; it is a 
purely random effect. The fact that slight defects are much
more common than severe defects is consistent with a
dispersion mechanism. An explanation based on mutations
is less satisfactory. It is true that severe forms
may affect the reproductibility rate and therefore the
transmission of possible genetic factors
but there would
be little difference in this respect between light forms
and very light forms.
\qbu Cleft lip and palate or more generally facial cleft. 
\qbu The positioning of the eyes (i.e. iris+pupil+lens)
also requires high accuracy because the two eyes
must move in a synchronized way. 
For each eye
positioning relies on two muscles (one on each side)
whose actions must be perfectly
coordinated. As it is not easy to achieve
such high accuracy requirements it is hardly surprising
that, as shown in Table 1 
strabismus is one
of the most frequent birth defects (2\% of births).
\qbu Heart valve defects are almost as frequent as
strabismus. More details will be given later.

\qA{Control procedures}

In industrial production there are control procedures all
along the supply and production chains. 
There are certainly
similar control procedures in the making
of living organisms. Although we do not know them
very well there has been progress in this direction
in recent decades. For instance, the role played by the
non-coding region of the genome (which represents 98.5\%)
is becoming clearer.
\qpar
 
Spontaneous abortion can be seen as a
control mechanism but the occurrence of live births with
severe malformations 
(e.g. anencephaly, that is to say newborns 
without a brain, whose prevalence is about 120 per 
million births) shows
that this control is insufficient.
It is true that apoptosis (that is to say programmed
cell death) is a local control mechanism, 
but it is surprising that massive defects at macro level
are not identified and corrected. In our industrial analogy
it would mean producing aircraft without wings%
\qfoot{It can be argued that this is an anthropocentric
view for indeed the ability to fly may not be the
main purpose. After all there are insects and birds
which have wings but cannot fly.}%
.
\qpar

The dispersion conception would also suggest more frequent
defects in highly complex organs than in simpler ones.
However, before we discuss this point we need to 
assess the reliability of defect statistics.

\qA{Why defect statistics give a biased picture}

The statistics of birth defects released by hospitals
give a picture which is biased in (at least) three respects.
\qee{1} Very serious defects usually will lead to early abortion
or still births. This fact can be illustrated by the following
data.  In 13,614 births that occurred in an 
hospital of Rajasthan (India) in 2012
there were 431 stillborn and
13,183 live-births. Among the stillborn, 18\% had a birth
defect whereas only 0.64\% of the live-births had a defect.
(Vyas 2016). Thus, many serious cases will not be included
if birth statistics are restricted to live-births.
\qee{2} Many slight defects will not be recorded
because they will give rise to symptoms only much later.
This can be the case even for heart defects; for instance
light valve defects or stenosis (i.e. narrowing)
will be noticed only at the age of 40 or 50.
It is the same problem for many other internal defects.
Whereas polydactily (i.e. more than 5 fingers) can
be detected visually just by inspection, many slight defects
of internal organs may never appear or appear 
only later in life.
\qee{3} For a complex organ like the brain, there is
no well defined border line between what is normal and
what is not. Thus, the fact that some persons can sing very
well while others cannot will not be considered as a congenital
defect. Even more serious defects (such as a propensity
to autism) will appear only
later on in life; as a result the respective 
role of genetic, environmental or dispersion factors will
remain unclear. For that reason, 
although the brain is by far the most complex organ of
a human body it will be left aside in the 
next subsection where we discuss the role
of complexity.

\qA{Are complex organs more affected by output dispersion?}

The manufacturing process of an airliner requires more accuracy
and controls
than the production of bicycles. Similarly, in a human body 
some organs are more complex than others. Obviously, the heart
is a more complicated device than the bones%
\qfoot{It is true that ``complicated'' has no obvious meaning.
Even a single cell is very ``complicated''. In addition 
it can be argued that the bone marrow is very essential.
What we mean here is that seen from outside
a pump (which is what the heart is) is more difficult to design
and build than a table leg.}%
,
the skin or even
the liver. Therefore, the fact that heart defects are the most
frequent congenital malformation comes as a nice confirmation 
of the dispersion conception. 
\qpar

In contrast, defects based on mutations are not expected to
follow the same
rule. It seems natural to admit that the number of mutations
(including harmful mutations)
is proportional to the number of genes involved
in the manufacturing of each specific organ. 
As each gene codes for a specific protein one would have
to admit that the number of proteins is in relation 
with the complexity of an organ.
If data are available such numbers could provide
a useful metric for estimating the complexity
of various organs.

\qA{The most frequent defects have close links with normality}

Defects, particularly minor defects, are usually ``in line'' 
with normal organs. In order to explain what we mean
by this expression let us consider polydactily defects.
Can the 6th finger appear anywhere?
\qpar

Firstly, one can observe that the additional finger
is never perpendicular to the hand.
Can it appear anywhere in the plane of the hand?
Observation shows that it is much more 
likely to appear on each side of the
hand (that is to say next to the thumb or little
finger) than next to the three inner fingers.
In other words the 6th finger is more likely 
to appear as an addition to the normal blue 
print rather than
as a drastic change in the normal design.
\qpar

A similar observation can be made for the heart valves.
Consider for instance the aortic valve which is located
at the beginning of the aortic artery. Whereas normally
it has three leaflets the defect which is by far the most
frequent is when two of them stick together.
The prevalence of this so-called bicuspid aortic valve (BAV)
defect is between 1\% and 2\%. In contrast,
the quadricuspid aortic valve (QAV) is a rare congenital 
anomaly with an incidence of only 0.01\% (Schaeffer et al. 2007).
\qpar

Why is the first defect more in line with the normal
valve than is the quadricuspid? 
The BAV originates from the fusion
of two existing leaflets whereas the QAV requires the creation
of an additional leaflet with corresponding changes to the three
others in order to make room for the new one. Such a defect would
require significant design changes.

\qA{Weak role of genetic factors in birth defects}

At first sight it may seem that the dispersion effect is only of
marginal importance compared to the genetic and environmental factors.
For a better assessment we use a methodology based on the
observation of pairs of twins.
\qpar

How similar are  monozygotic twins? 
The fact that they may look
``alike'' is not sufficient proof of their similarity.
This can be illustrated by a case reported in Williamson (1965, p.166).
In a study of family characteristics of congenital malformations
done in Southampton (UK) the author
reports the case of twins who were ``similar in hair color, eye color,
head shape, finger nail shape, teeth pattern and many other features''
but one of these twins was a hydrocephalic (too high pressure of
fluid in the brain)
male while the other was a normal male.
\qpar

It is true that no valid conclusion can be drawn from a single case
but this kind of observation is confirmed by a recent study of
6,752 monozygotic (MZ) twins and 13,310 dizygotic (DZ) twins
in California observed from
1957 to 1982 (Yu et al. 2019). 
\qpar

MZ twins share 100\% of their genome whereas DZ twins share on average
50\% of their genome (Yu et al. 2019, p.18).
In Appendix A we explain a
method for assessing the role of genetic factors.
When applied to the data given in Yu et al. (2019)
it leads (see Appendix A) to the conclusion that
genetic factors play 
in fact a fairly weak role in major congenital malformations.
This leaves free space for 
(i) environmental factors and (ii) for the 
dispersion effect described above.
\qpar

Is it possible to discriminate between (i) and (ii)?
For birth defects the only environmental 
factors which can play a role
are those which affect the mother. Many factors of that kind
were considered by researchers, e.g. age,
level of education, birthweight, birth order,
season of birth, smoking of the mother. 
It appears that only smoking of the mother%
\qfoot{In order to measure more accurately
the influence of this factor it would be useful to do
a comparative analysis covering a sample of countries with
highly different levels of tobacco consumption.}
is significantly associated
with congenital defects (Yu et al. 2019).
However, why should smoking of the mother affect one twin and
not the other?

\qA{Identification of output dispersion through phenotype variations}

Observation of uniparental reproduction offers a fairly direct
view of the effect of output dispersion. 
It allows the notion of ``pure line'' (also called ``inbred line''
or ``inbred strain'') 
to be defined in a rigorous
way as being formed by the offspring of a single individual.
In contrast, for sexual reproduction
a strain is considered 
inbred when it has undergone at least 20 successive endogenous matings
(brother-sister or parents-offspring) but 
even at this point the individuals are only nearly clones. 
That is why in the first half of the 20th century  there have been
many investigations of uniparental reproduction.
\qpar

Fig. 1 gives two illustrations. They are followed
by a table which lists causes of congenital anomalies.
%
%%-----------------------------------------------
%%%%   EXAMPLES DE DISPERSION DS LA DESCENDANCE
\begin{figure}[htb]
\centerline{\psfig{width=12cm,figure=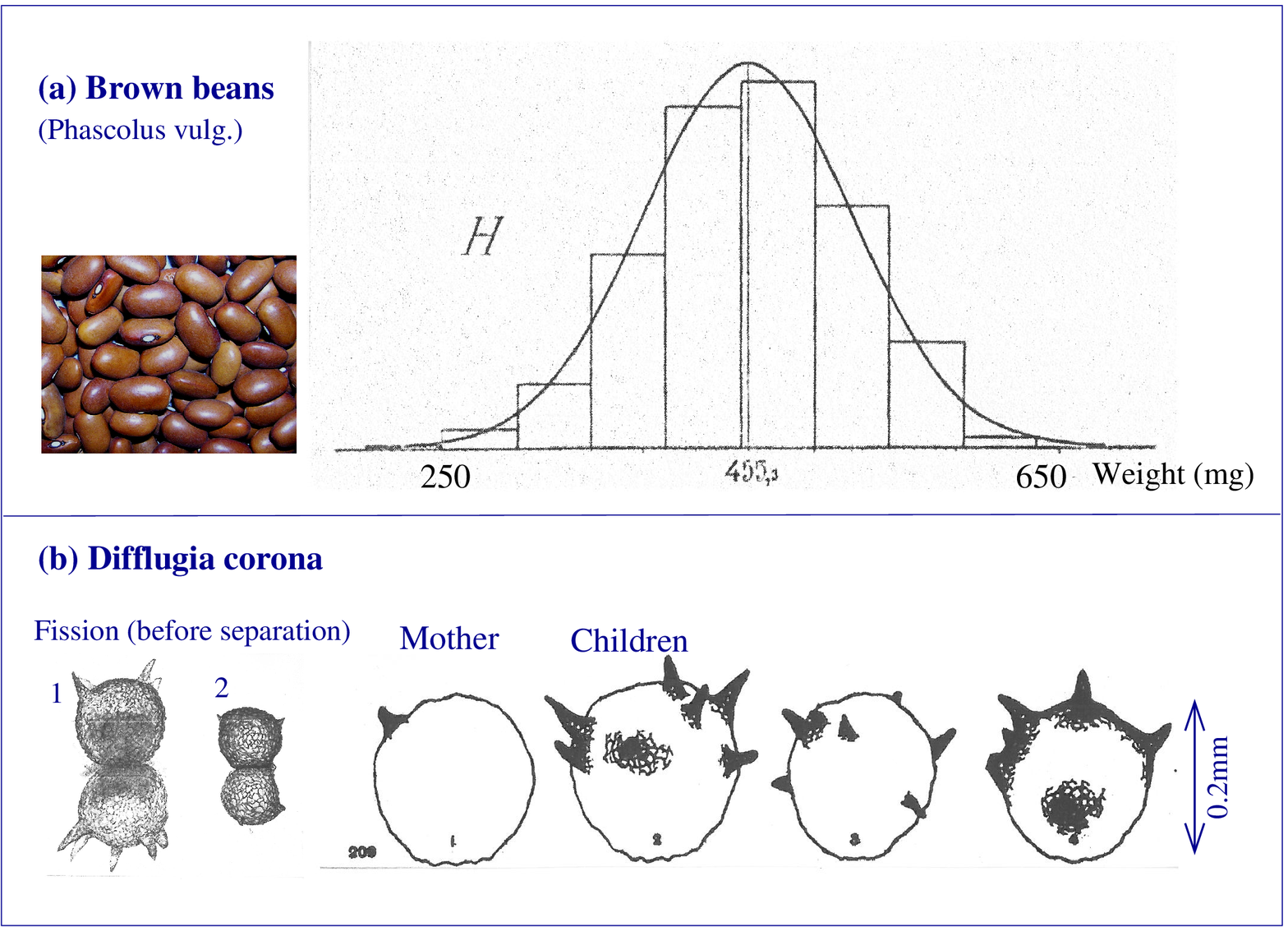}}
\qleg{Fig.\qhu 1\qhv Two examples of output dispersion
in uniparental reproduction.}
{{\bf \color{blue} Top:} 
Dispersion in the weight of 418 bean seeds in a pure line
obtained from a single grandmother seed through self pollination. 
The histogram is well described by a Gaussian distribution
of mean $ m=455 $mg and standard deviation $ \sigma=70 $mg which
gives a coefficient of variation CV$ =17.0\% $. For all nineteen
pure lines totaling 5,494 beans CV$ =19.9\% $.
These experiments
were done by Wilhelm Johannsen in 1900-1902.
{\bf \color{blue} Bottom:} Dispersion in the aspect of {\it Difflugia corona},
an unicellular protozoan living in water. As reproduction is
by fission (as shown on the left-hand side for two pairs
differing in size) 
all 3 descendants of the
first individual (at the left) are clones. However, there
are variations in their aspect at time of fission, particularly 
in number of spines on the shell. Note that natural self-pollination
is not exactly the same thing as asexual
uniparental reproduction; the later produces real clones
whereas in the former (when performed naturally)
there is a high degree of inbreeding
which however may be somewhat less than 100\%.} 
{Sources: Johannsen 1903 (p.22-28), Jennings 1916 (p.438-439).}
\end{figure}
%-------------------------------------------------

The main difference between the two experiments 
shown in Fig.1 lies in the
number of successive generations that can be observed.
For Johannsen's beans there was only one harvest per year 
whereas under good conditions the protozoans reproduced
at intervals of 3 to 5 days, that is to say almost one hundred
times faster than the beans. Another difference is that the
second experiment relied mainly on results expressed in integers:
either the number of spines whose range is 0-7 or the number
of teeth around the mouth% 
\qfoot{The mouth cannot be seen on the picture describing the
fission process because it is located at the separation
between the mother and daughter cells.} 
which is an integer smaller than 17.
\qpar

A study with a similar objective was published in 1915
by Ms. Ruth Stocking which was based on variations occurring
in paramecia ({\it Paramecium caudatum}), a large unicellular
organism which lives in fresh water. Here again, as reproduction
is by fission (and does not involve conjugation episodes),
the descendants of each single individual will constitute a pure line.
The study focused on the shape of the paramecia. 
A recapitulation figure (p. 408) shows a bewildering diversity
of forms from the standard ellipse to strange shapes with many
tentacles. 
\qpar

%
%%-----------------------------------------------
\begin{table}[htb]

\small
\centerline{\bf Table 2: Mechanisms related to congenital
anomalies.}

\vskip 5mm
\hrule
\vskip 0.7mm
\hrule
\vskip 0.5mm
$$ \matrix{
&\hbox{Mechanism}\hfill & \hbox{Passed} & 
\hbox{Identification test} \hfill & \hbox{Example}\hfill \cr
&\hbox{}\hfill & \hbox{to} & 
\hbox{} \hfill & \hbox{}\hfill \cr
&\hbox{}\hfill & \hbox{offspring} & 
\hbox{} \hfill & \hbox{}\hfill \cr
\qtb
&\hbox{}\hfill & \hbox{Yes/No} & 
\hbox{} \hfill & \hbox{}\hfill \cr
\noalign{\hrule}
\qth
 &\hbox{\color{blue} Design glitch}\hfill & \hbox{} & 
\hbox{} \hfill & \hbox{}\hfill \cr
1 &\hbox{Mutation in DNA of gem cells}\hfill & \hbox{Yes} & 
\hbox{Genome sequencing} \hfill & \hbox{Trisomy 21}\hfill \cr
2 &\hbox{Mutation in DNA of somatic cells}\hfill & \hbox{No} & 
 \hbox{Non inheritable} \hfill & \hbox{Cancer}\hfill \cr
 &\hbox{}\hfill & \hbox{} & 
\hbox{abnormal cells} \hfill & \hbox{}\hfill \cr
 &\hbox{\color{blue} Manufacturing glitch}\hfill & \hbox{} & 
\hbox{} \hfill & \hbox{}\hfill \cr
3 &\hbox{Environmental interference}\hfill & \hbox{No} & 
\hbox{Epidemiological studies} \hfill & \hbox{Effect of nicotine}\hfill \cr
4 &\hbox{Random output dispersion}\hfill & \hbox{No} & 
\hbox{Uniparental reproduction} \hfill & \hbox{Strabism}\hfill \cr
 &\hbox{\color{blue} }\hfill & \hbox{} & 
\hbox{} \hfill & \hbox{}\hfill \cr
 &\hbox{\color{blue} Repair mechanism}\hfill & \hbox{} & 
\hbox{} \hfill & \hbox{}\hfill \cr
\qtb
5 &\hbox{Apoptosis (programmed cell death)}\hfill & \hbox{No} & 
\hbox{} \hfill & \hbox{Finger separation in embryo}\hfill \cr
\noalign{\hrule}
} $$
\vskip 1.5mm
Notes: Four comments are in order.
\qbu It is the word ``random'' which characterizes the
difference between items 3 and 4. It
means that dispersion in outputs occurs
even in optimum conditions, i.e. when no 
harmful environmental factor is present.
\qbu Mutation and repair mechanisms can hardly be separated
for most often we can see only their combined effects.
If the cells resulting from a somatic mutation are 
quickly eliminated through apoptosis nothing will appear.
\qbu Uniparental inheritance tests allow
a distinction between (2) and (3)+(4). If, as seems natural,
the amount of somatic mutations increases with time,
their contribution to {\it congenital} anomalies should be fairly
small. Moreover, when (3) can be excluded in
the controlled environment of a laboratory experiments, then 
(4) seems the most likely mechanism for
the abnormalities shown in the text.
\qbu Epigenetic mutation was not included in the table for
its status does not seem clearly defined. For instance,
one of its mechanisms involves
the addition of methyl radicals $ \hbox{CH}_3 $ to the molecules
composing the DNA but what triggers this addition remains unclear.
%Sources: 
\vskip 2mm
\hrule
\vskip 0.7mm
\hrule
\end{table}
%%-----------------------------------------------

How can one account for the variations observed in 
those experiments? Standard factors are listed in Table 2.
Item 1 is clearly excluded because the changes were not
inheritable. Item 2 seems unlikely.
If somatic mutations are random and independent from one another
their number {\it must} be proportional to the number of cells%
\qfoot{This statement just results from basic probability theory.
Peto's paradox relies on what happens not at cell level
but at the level of the organism (``Why don't all whales
have cancer?''). A mutated cell will lead to cancer only
if it is not removed by the immune system. Is the immune system
of the mice used in laboratories not affected by the fact that
they are pure line mice?} 
and to the time interval. Thus, for unicellular organisms observed at
fission time this effect should be minimal.
\qpar

What can be said about item 3?
With a little imagination one can easily suggest possible
environmental factors. Thus, for beans one can mention the 
position of the beans in the pods and the location of the pods
on the plant. However, why should such discontinuous factors
lead to almost perfect Gaussian distributions?
For the protozoa which were raised in 
laboratory conditions and identical medium it is
more difficult (yet not impossible) to cite environmental factors.
In a general way, however,
in order to make a convincing case for a specific
environmental factor, evidence must be provided
showing that in a series of tests it has indeed
the claimed effect. Otherwise it would be just an {\it ad hoc}
explanation.
\qpar

It is surprising that item 4 is almost never mentioned.
In particular, we did not find it in the 
numerous papers of the 1910s and 1920s analyzing asexual
reproduction. Yet, is it not a natural mechanism?
It can easily account for continuous variability
as described in Johannsen's paper because its randomness
leads naturally to Gaussian distributions. 
Through the Central Limit theorem of probability
the occurrence of a random discrepancy
$ X_i $ at each step $ i $ of a multi-step pathway gives
a nearly Gaussian distribution for the sum of the $ X_i $
(at least if the $ X_i $ are independent).\qL
Through the
hole and shaft  mechanism described below item 4 can also account
for variability by leaps, as happens for spine numbers
or a similar effect for tentacle numbers in Lashley (1915).
\qpar

In principle if the manufacturing process is known
it should be possible to compute and predict the
variability of the output (except if butterfly effects
play a major role).
In other words, this framework
can really be tested. Although 
in the present paper we limited ourselves to 
qualitative or semi-quantitative tests, subsequently
it should be possible to find cases simple enough to
allow modeling.

\qA{Outline of the paper}

The paper proceeds through the following steps.
\qee{1} First, we explain why random output fluctuations
are inevitable in any production process. It is only thanks
to a sound management of defects that an assemblage of
several (defective) parts can be made workable.
Depending on the specific industry, those management systems
use different ways. We will focus on the tolerance system in
use for mechanical systems because it is probably
the easiest to understand.
\qee{2} Secondly, we explain in what respects the two
phases of human mortality, the ``wear-in'' and
``wear-out'' phases, bear close resemblance
with the failure modes defined in reliability
engineering.
\qee{3} If simple technical devices can give us 
a better understanding of how to achieve minimal manufacturing
defects, is it not natural to try the same
approach for living systems? For instance,
is the shape of the age-dependent
infant mortality of simple living systems similar
to or different from that of humans? This leads us
in our conclusion to
outline an agenda of cross-species investigations.

\qI{Fault-tolerant design}

In order to make industrial production able to cope with 
output discrepancies in the supply chain
appropriate systems have been developed.
In the following subsection we explain
briefly the tolerance system for mechanical devices.
In recent decades
much attention has also been given to electronic
semiconductor systems because of the high complexity reached by
such systems which may 
have millions (or even billions) of components (Dubrova 2013).
In a broad way, the purpose is always the same and can 
well be summarized
by the title of a paper written by John Von Neumann in 1952, namely:
``Synthesis of reliable organisms from unreliable components''
(Von Neumann 1956).

\qA{Tolerance issues in the industrial production of mechanical devices}

First of all, it should be realized that 
mechanical operations involve
inherent output variations. This was already mentioned earlier
in an informal way; let us see more precisely how the 
tolerance system can deal with it.
\qpar

Two holes made on a lathe with the same drill bit
(say of 10mm diameter)
in an aluminum cylinder will
in fact not have the same diameter. 
The boring operation will introduce
a small but unavoidable random error. 
For instance,
the diameter of the holes may be $ 10.003 $mm and
$ 9.996 $mm respectively; 
naturally, the measurement  
introduces an additional uncertainty
which will be ignored here for the sake of simplicity.
\qpar

One may think that this small difference is
of little importance
but suppose that this hole is destined to receive
a shaft which has a diameter of $ 10.000 $mm. This will
be possible for hole 1 but not for hole 2. In short,
even small discrepancies may prevent assemblage.
\qpar

As already mentioned,
in embryo-genesis there is a somewhat similar problem
when two separate sheets are expected to join.
In such cases even a small discrepancy in growth
velocities may disrupt normal closure. This
may create a defect of the neural tube
which results in a birth abnormality called ``Spina bifida'',
a Latin expression
which means ``spine split in two''. 
Similarly, disruption of the closure of
the left and right facial sheets may result
in what is called a cleft lip and cleft palate.
We come back to this point below.
\qpar

A related case is the genesis of the furcula.
In humans the furcula consists of two separate bones
called clavicles or collarbones. On the contrary in birds it
is a single V shaped bone called furcula (latin for small fourk)
or wishbone.
Located in the upper chest of birds it
is an essential structural element which allows them
to move their wings; it also acts as a mechanical 
spring during flight.
On day 13 of the 21-day long embryogenesis
of chicks 
the left and right collar bones meet and close together to
form the furcula. It can be predicted that 
even  small discrepancies can
prevent good working of this critical element.

\qA{The tolerance system as a way to mitigate the effects of
manufacturing defects}

Mechanical engineers have developed
a system of standardized {\it tolerances}. 
In this context a 
tolerance is a specification which gives not only
the nominal dimension but also the allowed margin.
As an example, for the previous hole, the specification
would be: 10 +0.015−-0\ mm,
meaning that it may be up to 0.015 mm larger than the
nominal dimension, but 0 mm smaller (that is to say
it should not be smaller than 10mm). 
\qpar

The task of the engineer is to
give for every dimensions
appropriate tolerances so that, if 
respected, the device will work. 
For each separate
part the technician who makes it will check whether
or not it is ``within tolerances''. If it is not,
it will be discarded and replaced by a suitable one.
\qpar
%
%%-----------------------------------------------
%%%%   TOLERANCE - DEFECTS
\begin{figure}[htb]
\centerline{\psfig{width=15cm,figure=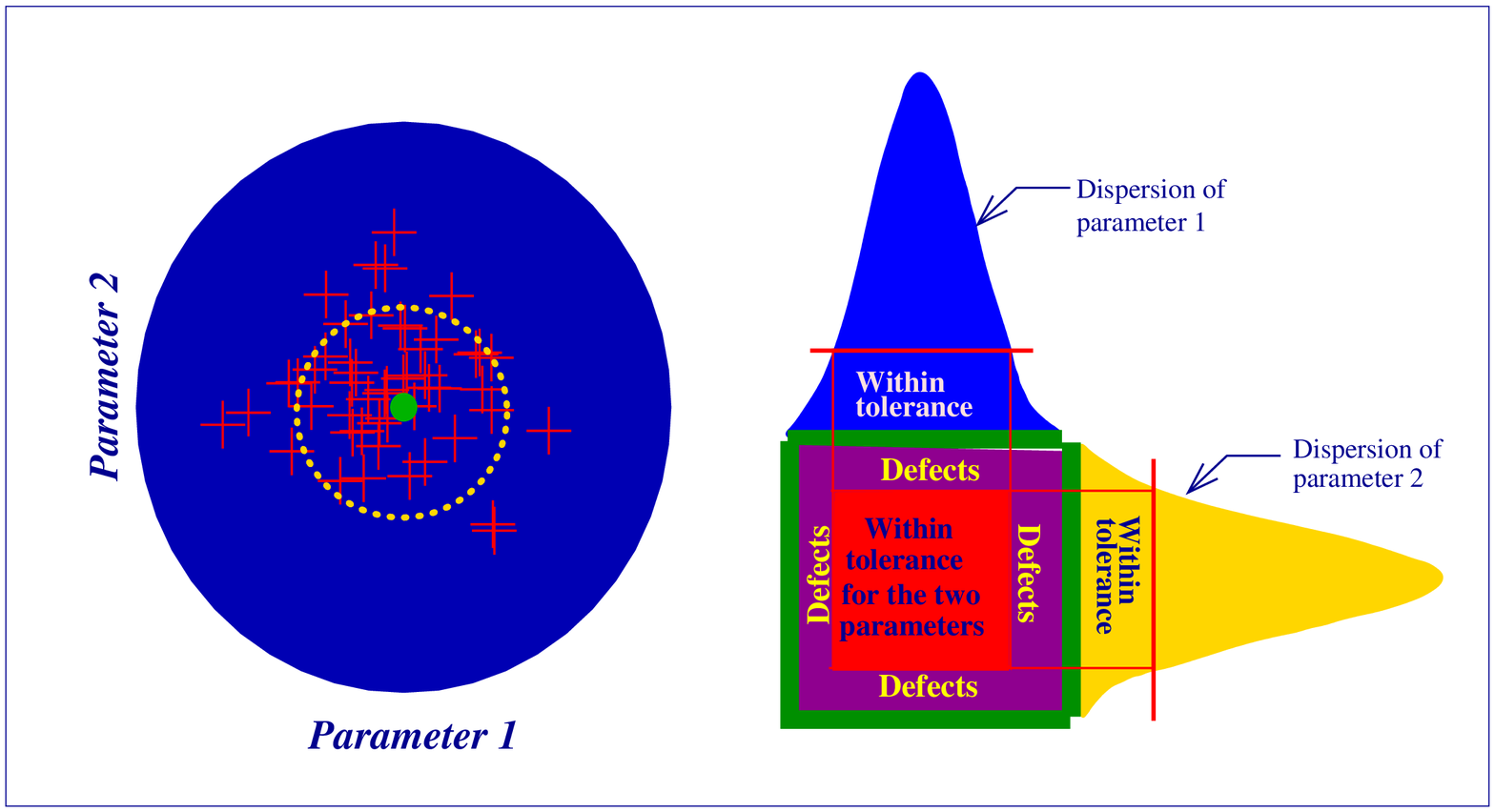}}
\qleg{Fig.\qhu 2\qhv Within and out of tolerance areas when a process
depends on two parameters.}
{In this schematic representation it is assumed that
a process depends simultaneously on two parameters,
each of which has a Gaussian distribution. The green dot represents
the (optimal) design values of each parameter. As
illustrations one can mention the following cases. (i) The green
dot corresponds to the ideal center of a hole that is
drilled into an aluminum cylinder. Actual centers 
in 50 successive realizations are represented
by the red crosses. Although never exactly at the design location, the
effective centers may be close to it and fall within the tolerance
domain represented in yellow (note that it may have
another shape than a simple disk).
(ii) For a chemical reaction
parameter 1 may be the concentration of one component and parameter 2 the
concentration of the other. Then, the green dot corresponds to the optimum
concentrations.
For the process to unfold successfully both parameters
must be within tolerance which means that all cases
which fall in the magenta region will not work well and may lead 
to defects. For a process which has more than two parameters
the acceptable zone would be reduced even further. 
Such additional parameters
could be for instance the temperature and pH.}
{}
\end{figure}
%-------------------------------------------------

There are similar tolerance systems for electrical
elements such as capacitors or resistances. The
specification (often written on the element itself)
may indicate the nominal value (e.g. 100$ \Omega $), the
margin of error (e.g. $ \pm 1\% $), 
the temperature range (e.g. 5 to 35 Celsius degrees). 
\qpar

One could summarize the specification
procedure by saying that the
science of engineering is to make working devices 
with spare parts which, strictly speaking, are all
defective in the sense that their values
differ from the nominal values (but are within tolerance
margins).
This mechanical example is useful because it allows
a clear understanding of the problem but since living organisms
are not made with nuts and bolts, nor
with resistors, one must explain how this
should be adapted.

\qA{Output dispersion in biological systems}

At first sight one may think that the two cells
produced in the fission of a parent cell
are exactly identical. The previous discussion suggests
that in fact they are not, but does not explain the why and how.
Basically, biological processes consist in a succession
of physico-chemical reactions.
In order to give an intuitive feeling of why
such reactions are sensitive even to fairly small
condition changes we will make three points. 
(i) First we emphasize the relatively high frequency of errors
in protein folding. (ii) Secondly, we explain how spatial 
factors play a great role in reactions involving enzymes.
(iii) Thirdly, we consider a simple reaction whose 
high sensitivity to temperature may be familiar to many
readers.
\qee{1} It has been recognized that 
``errors arise at all steps of protein synthesis, from transcription to
protein folding, and have
widespread phenotypic consequences''. Due particularly
to the ``fragility'' of protein folding mechanisms ``errors in
protein synthesis are orders of magnitude more
frequent than DNA-replication errors''
(Drummond et al. 2009). This review paper contains a table 
which lists a number of errors along with their estimated frequency.
\qee{2} One hallmark of the present paper is to
emphasize the role of geometrical and positional factors.
Here is another case of that kind.
We know that enzymes (most enzymes are special kinds of proteins)
act as catalysts of chemical reactions.
In fact, they are highly sophisticated catalysts in the sense
that they can play this role not only for one specific reaction
but for several. In addition their activity can be modulated according
to needs. In other words, they are a kind of multipurpose control
station, somehow like the control room of a power plant.
The multipurpose capability comes from the fact that at their
surface they have several so-called active sites where the reaction
will take place; each active site is coupled with a so-called
allosteric (meaning ``other place'')
site  which will bind with  control molecules that can be either
activators or inhibitors. Needless to say, 
if a control molecule is attached near but 
somewhat off the right location
its regulation function will not be well implemented. 
With allosteric sites that are particularly cramped
there can be situations similar to the hole and shaft 
case where even a small shift can greatly affect the enzyme
and therefore the reaction that it is supposed to catalyze. 
To make things even more complicated, one should add
that many enzymes do not work well if they are not
bound to helper molecules called cofactors. 
\qee{3} Our third illustration is a process which 
may be familiar to many readers.
As is well known, a mayonnaise
is made by slowly adding oil to egg yolk, while whisking
vigorously with a fork.
An emulsion will form made of small oil droplets. 
These droplets are strongly held together by van der Waals
intermolecular attraction forces which cause
the high viscosity of mayonnaise
(Depree et al. 2001). Addition of
mustard contributes to the taste and further stabilizes the emulsion.
\qL
This, at least, is the theory. 
\qL
In fact, the operation may
fail (i.e. no emulsion forms) for various reasons.
\qee{i} It fails when the oil is added too quickly.
\qee{ii} It fails when the temperature of the oil is too high;
as a matter of fact,
it works best when the oil and egg come directly from
the refrigerator. 
\qee{iii} Another reason for failure may be the presence 
on the fork of traces of a product
which prevents the formation of the emulsion.
\qL
In short, we have here a simple physico-chemical
process which has fairly strict tolerance 
specifications. If two or several processes are involved either
successively or at the same time, the tolerance area is
further reduced (Fig.2).

\qI{Salient features of embryonic mortality}

In previous sections it was suggested that a manufacturing
process which involves major innovations is more prone
to faults than mere cell reproduction by fission.
That is why, for instance, the transition from stem cells
to fully differentiated cells is a more
challenging task than duplication.
\qpar

The process by far the most innovative
is the transition from a zygote, i.e. a fertilized cell,
to a fully developed embryo. Within a fairly short fraction
of the order of 10\% of the embryonic period,
a completely new organism will be created and each step 
is highly dependent upon the satisfactory outcome of
previous steps. In other words, this is a critical
development process in which major faults are expected to occur
with significant probability.

\qA{Implication of geometrical abnormalities for development
of the embryo}

Fig.3a shows  {\it position
anomalies} occurring
in the early steps of embryogenesis and Fig.3b indicates that they
have adverse implications as revealed by the fall in hatching rates.

%
%%-----------------------------------------------
%%%%   MALFORMATIONS DS LES PREMIERS STADES DE L'EMBRYON
%%%%   
\begin{figure}[htb]
\centerline{\psfig{width=17cm,figure=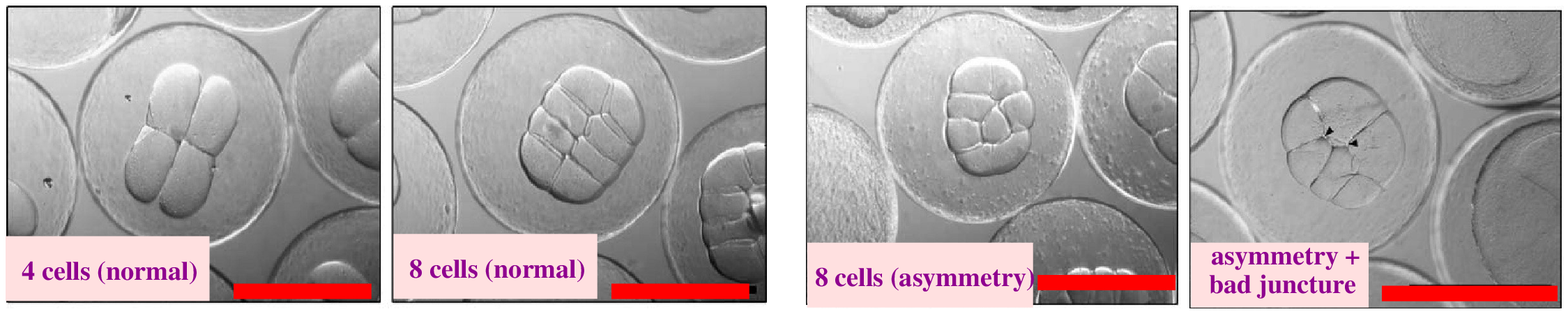}}
\qleg{Fig.\qhu 3a\qhv Cleavage abnormalities in haddock embryos.}
{Examples of abnormalities occurring in the first steps
of embryogenesis. The growth process starts with one fertilized
cell, then subsequent steps every 20mn with 2,4,8,16,$ \ldots $ cells.
The pictures show that early defects can occur already in
the 8-cell step. Normal development is shown on the left-hand side
and abnormal development on the right-hand side.
The red segment corresponds to 1mm.
Apart from the two cases shown here
three other sorts of abnormalities are described
in the same paper, namely (i) unequal sizes of the cells
(ii) cellular outcrops where one or two cells protrude
from the main group of cells. (iii) Separation of the 8 cells
into two disconnected sets.
In the following figure it is
shown that such abnormalities result in lower
hatching rates that is to say in increased embryonic mortality.}
{Source: Adapted from Rideout et al. (2004,p.219)
}
\end{figure}
%-------------------------------------------------

%
%%-----------------------------------------------
%%%%   EFFET DES MALFORMATIONS SUR LE TX D'ECLOSION
%%%%   
\begin{figure}[htb]
\centerline{\psfig{width=12cm,figure=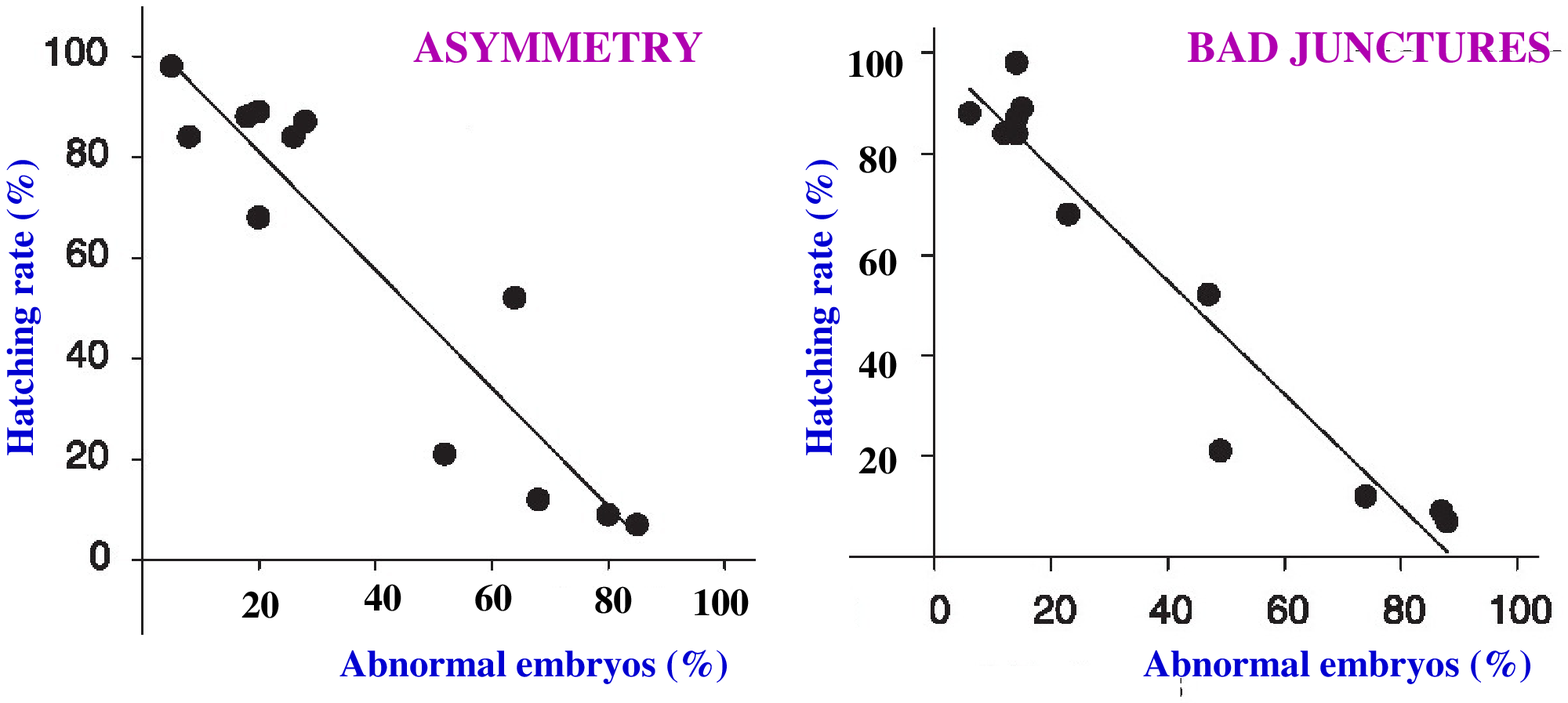}}
\qleg{Fig.\qhu 3b\qhv Hatching rates for embryos involving
malformations.}
{Hatching rates were measured for 12 samples containing
various proportions of defective embryos. The coefficients
of linear correlation are equal to $ r=0.93 $ and $ r=0.96 $,
respectively. Similar correlations are obtained for
size and outcrop anomalies.}
{Source: Adapted from Rideout et al. (2004, p.222)
}
\end{figure}
%-------------------------------------------------
%
\qA{Age-dependent embryonic mortality}

In demography age-specific death rates are a key-variable%
\qfoot{From a physical perspective the resolution of
demographic phenomena into age-specific components 
is similar to
frequency analysis of physical phenomena; for
more details see Berrut et al. (2017).}%
.
%
%%-----------------------------------------------
%%%%   MORTALITE EMBRYONNAIRE
%%%%   CALCULS FAITS DS: MAPY#EMBRYON
\begin{figure}[htb]
\centerline{\psfig{width=10cm,figure=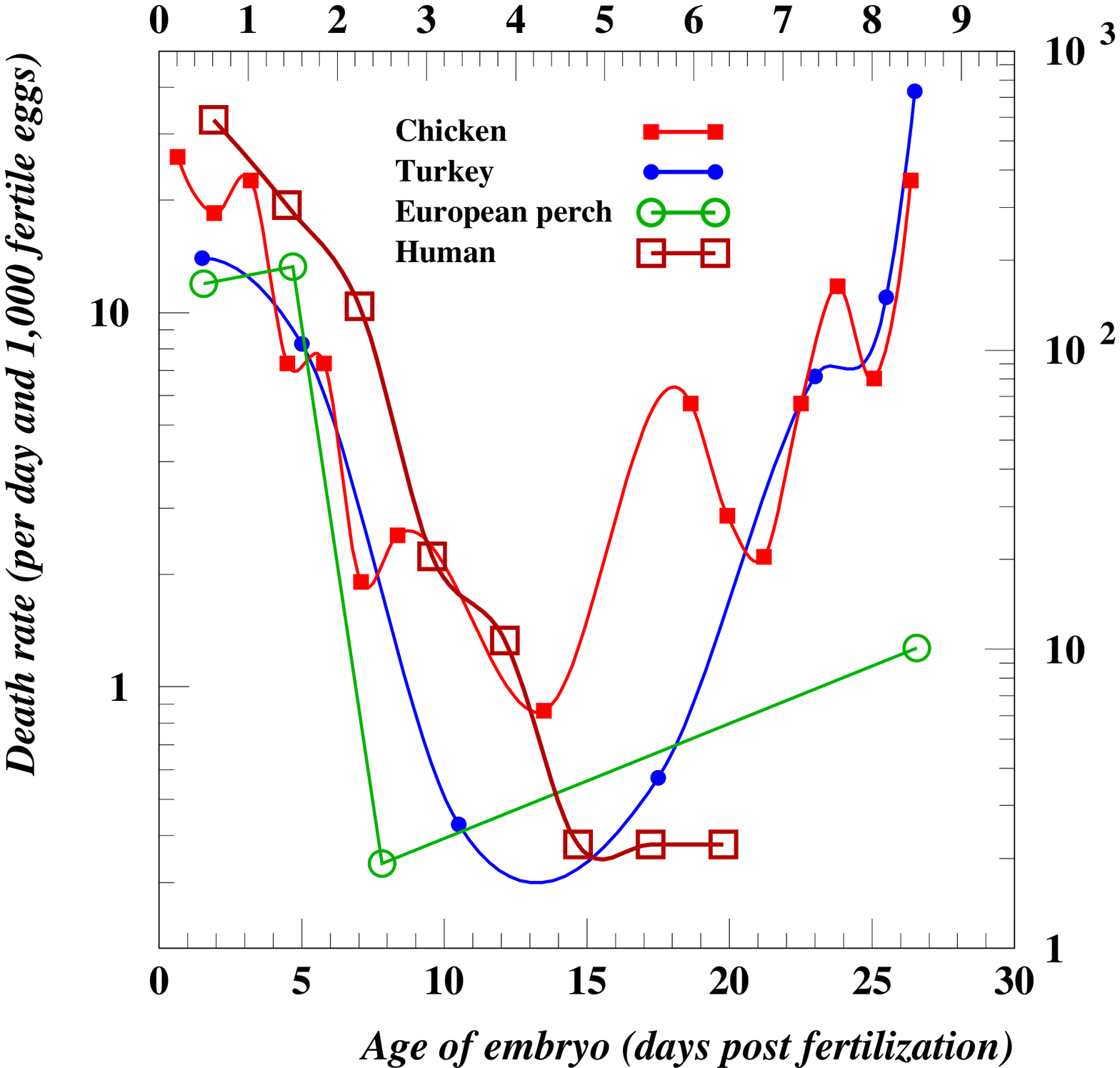}}
\qleg{Fig.\qhu 3c\qhv Embryonic mortality rates.}
{The bottom and left-hand side scales are for birds.
The top and right-hand side scales are for fish
(ages are also expressed in days).
The scales for humans (not shown) are as
follows: the age scale starts
with the 4-7 weeks gestational age interval and ends at 50 weeks.
The vertical scale (expressed in rates per 1,000 pregnancies)
starts at 2 and ends at 150.
Note that the data provided by vital statistics agencies
usually start only at 20 weeks. The present data for the intervals
between 4 and 20 weeks
were obtained through a special study covering a 4-year period
(1953-1956). Note that the age scale of the chicken case has
been extended from 21 to 24 days to facilitate the comparison
with the turkey case. Note that the perch curve is made
of straight lines because there are too few data points
to use a smoothing option.}
{Sources of the data: Chicken (broiler): Pe\~nuela et al. (2018, p.6505),
number of fertilized eggs $ ( n )\ = 3,146 $;
turkey: Fairchild et al. (2002, p. 262), $ n=51,764 $; 
European perch: 
Alix (2016, p. 161), $ n=13,500 $;
humans: French et al. (1962, p.840,844), $ n=3,083 $. 
}
\end{figure}
%-------------------------------------------------
%
In the embryonic phase they are paralleled by
mortality rates as a function of post-fertilization age 
which, therefore, should also be seen as a key-variable.
Curiously,
it attract little attention so far; as a result,
such data are available for only few species. Fig. 3c
presents data obtained by three high-accuracy studies
for bird and fish species. The graph also shows human data,
albeit with the drawback of starting 4 weeks after conception.

\qA{General observations about embryonic mortality}

What can be said about the role of mutations 
and environmental factors? \qL
For the animal experiments described in Fig.3c all embryos
were raised in identical conditions so that exogenous
factors can hardly explain why some embryos are affected
by severe anomalies while others are not.
\qpar

Mutations are certainly responsible for some anomalies
but under the assumption of a uniform mutation rate
it seems difficult to explain the huge changes affecting
the death rate. For turkey or chicken eggs why
should there be more lethal
mutations on day 1 than on day 11?
\qpar

For all four species, there is a sharp fall of the death rate
between fertilization and
the subsequent leveling off. For the turkey, perch and human cases 
the death rate is divided by a factor
of about one hundred whereas for chicken the factor
is about 30. However this last factor is affected by a 
substantial uncertainty
because of the small numbers of deaths; indeed, 
between days 8 and 14 
the daily death numbers are all smaller than 6
with three of them being zero or one%
\qfoot{This is in spite of the fact that the experiment 
involved 3,240 eggs and that 471 of these embryos died.
As the turkey experiment involved 10 times more eggs
its results are more reliable.}%
.  
\qpar

The fact that for the avian cases there is a second peak
on the right-hand side whereas no similar peak appears
in the two other cases is due to the fact that birds
have to pierce the shell of their eggs which is a difficult
task. If early neonatal death rates would be included into
the embryonic phase there would also be a left-hand side peak
in the perch and human cases. In other words, this difference
is related to how one defines the end of embryogenesis.
\qpar

As a last point we wish to compare the absolute magnitude of the 
death rates at the beginning of the embryo development.
For this comparison we leave apart the human case for reasons 
which are explained below.
\qpar

For turkeys the first data point which is an average for the first three
days stands at 14 per day and per 1,000 fertile eggs.
For chicken the average for the first three days stands at 22 which is
close.
\qpar

For the European perch the data point for the first day stands
at 167 that is to say about 10 times higher than for the birds
in 3 days.
The interpretation of this difference remains
an open question at this point.

\qA{Avian species}

To the two avian cases shown in Fig. 3c one can add that a
similar pattern was observed for several other avian species,
e.g. pigeons, 
doves, ducks, grouse, pheasants and quail (Romanoff 1949).
\qpar
 
The fact that some of these deaths are due to fairly random
conditions can be illustrated by the case of
malpositions. It has been observed that one half of all
chick embryos which die between day 18 and 20 were in
abnormal positions (Hutt 1929). In order to understand the reason
one should recall that the lungs of chicks start to work
shortly before they begin to break the shell of the egg.
However, to make that possible they must have access to the 
air cell which is on the blunt tip of the egg. If for some
reason their head cannot move in time  to the right location
the chicks will die. Moreover, to pierce the eggshell is 
quite a challenge%
\qfoot{For that purpose the chick is using a special
``tool'' in the form of a so-called egg-tooth which is a sharp
temporary structure on the top
of the beak. There is also a special ``hatching muscle''
which serves the purpose of activating the egg-tooth.}%
.
If, for some reason, the eggshell is too hard or too thick
the chick may be unable to break it.

\qA{Fish species}

The embryonic phase of fish can be studied easily due to the 
fact that the fertilization of the eggs occurs outside of the
body of the female.
For that reason one can get reliable death data
even for the very early part of the cycle. For instance, for
zebrafish as the first division of the fertilized embryo occurs
less than an hour after fertilization one should be able to
get hourly death rates.
Unfortunately, such investigations did not attract much attention
so far. To our best knowledge the case of the European perch described
in Fig.3c is an unparalleled study of fish embryonic mortality.  

\qA{Human fetal deaths}

The study described in French et al. (1962) took place
in the island of Kauai in the state of Hawaii. During the
four years of the study there were 3,083 pregnancies, 273
fetal deaths and 2,777 live births. These are of course
small numbers due to the fact that the island's population
was only 30,000. The reason for doing the study in this
place was the existence of a well organized
network of medical personnel. 
\qpar

Very early fetal deaths can only be noticed by the
women themselves. That is why this part of the death rate
curve must be recorded through special surveys involving
a devoted network of physicians and medical personnel. 
Standard fetal death statistics as provided
by hospitals include only pregnancies which lasted more
than 20 weeks. 
\qpar

In the three other cases of Fig.3c the procedure was to
observe a sample of $ N $ eggs in the course of time
and for each subsequent day to record the number of surviving
embryos. Clearly, it was not possible to use the same procedure
here. As pregnancies and fetal deaths were recorded
in a continuous way the whole process required more
intricate and less transparent computations. 

\qA{How can one explain that the death rate is highest
at the beginning of embryogenesis?}

Here is a tentative interpretation of
the fact observed in Fig.3c that the 
death rate is highest on the first day of the embryogenesis.
\qpar

In principle the organism of the mother produces
embryos equipped with all that they need to grow.
But, as for any real process, there are necessarily
faults and defects.
The embryos in which some important ingredients
are missing will be unable to grow and instead will die.
As these faulty embryo are gradually eliminated the death
rate will decrease just as observed.
\qpar

At present this mechanism is purely speculative but
the interesting point is that it can be tested.
How? \qL
Consider for instance the case of zebrafish embryos.
Two hours after fertilization the embryo has about 64 cells.
If the embryo is able to reach this point it means
that it is well equipped, at least for the cleavage phase.
In contrast, one would expect the faulty embryos to be eliminated
very shortly after the beginning of the embryogenesis.
This means that the death rate should be highestin the very first hours. 
In other words, this explanation can be tested by measuring
the embryonic death rate every 2 or 3 hours during the
first 24 hours.

\qA{A conjecture about embryogenesis in unicellular organisms}

In unicellular organisms is there a process 
similar to embryogenesis which precedes the birth
of a new organism? Formally no, but functionally yes.
For instance in the prokaryotic 
bacterium {\it Caulobacter crescentus}
the initiation of replication starts some 2 hours 
before division actually occurs (Laub et al. 2000,p.2145).
This phase 
(which consists of successive so-called 
G$ _1 $, S and G$ _2 $ transitions) 
can be considered
as a kind of embryogenesis during which the new organism
is made ready for autonomous survival.
\qpar

Naturally, the success rate of the complex transformations
which take place is certainly not 100.00\%. 
For instance, it has been shown (Cryms et al. 1999) 
that hyper expression of
one gene (named {\it podJ})
involved in a crucial transition at the beginning
of the replication process causes a lethal cell division defect.
Thus, it is conceivable that random fluctuations in the
concentration of this protein will lead to a percentage
of failures.\qL
This means that, in the same way
as there is an embryonic death rate, there will be 
a predivisional death rate. The magnitude of this death rate
will give an estimate of the sensitivity of the process to
random variability. The more sharp requirements are included
in the design of the process, the higher the expected failure rate.
\qpar

In the wake of the division, as indeed in a more general way 
after any major transition,
one expects a phase of infant mortality
during which the death rate of the daughter organisms
will start from an inflated level and then
decrease as the screening progresses. Those organisms 
for which the replication process has
been carried out to its end but which nevertheless 
are not completely
fit for an autonomous existence, will die.
\qpar

\qA{Influence of temperature on hatching rate}

Fig.3d shows a striking influence of temperature on
the average mortality rate during the 21-day long of the
embryonic phase of chicks. 
In terms of hatching rate which is perhaps more suggestive
(but less apropriate for cross-species comparison)
there is an increase from 10|5 at 35.8 degrees to 88\%
at 38.1 degrees and then a fall to 50\% at 39.8 degrees.

%
%%-----------------------------------------------
%%%%   MORTALITE EMBRYONNAIRE POUR POUSSINS
%%%%   CALCULS FAITS DS: MAPZ#CHICKS
\begin{figure}[htb]
\centerline{\psfig{width=10cm,figure=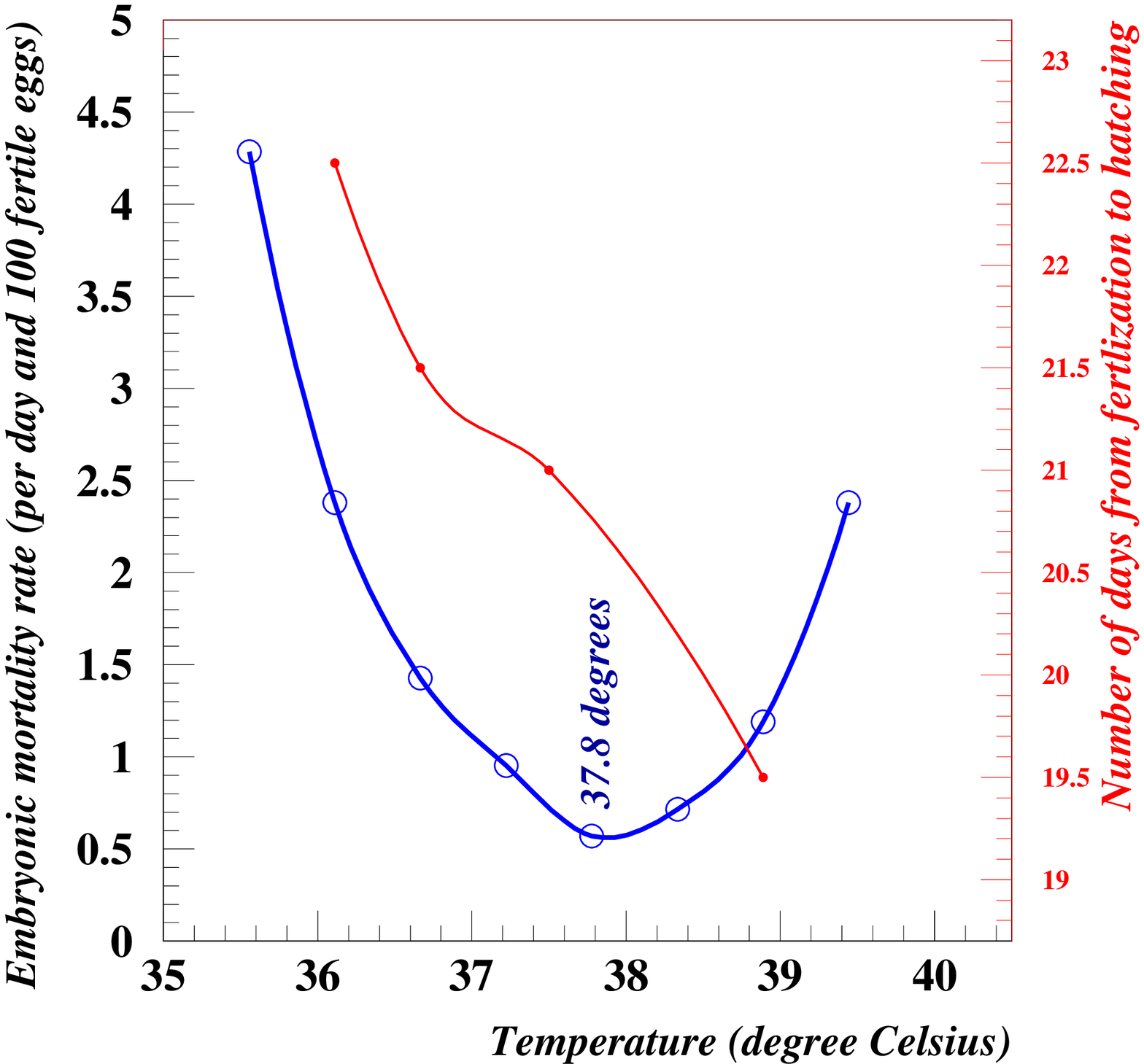}}
\qleg{Fig.\qhu 3d\qhv Average embryonic mortality rate for chicks
from fertilization to hatching.}
{Apart from the mortality rate, the graph shows also
the length of the embryonic phase.
On account of the fact that the speed of most chemical reactions
increases with temperature one is not surprised by the shortening
of the embryonic phase. On the contrary,
the fact that the mortality rate exhibits
a sharp minimum requires an explanation. A good test of our
understanding of this manufacturing process will be our ability
to predict (at least approximately) the optimal temperature.}
{Source of the data: ISA (2009).}
\end{figure}
%-------------------------------------------------

It can of course be argued that the temperature is an
environmental parameter but this is just a label and would not help
to explain the behavior seen in Fig.3d. It is clear that 
it is only through a better understanding of the manufacturing
process that we can hope to predict the shape of the mortality curve;
needless to say, the temperature is an essential variable
in this process.

\qI{Salient features of the two phases of human mortality}

Our main goal in this section is to show that the curves of
age-specific infant mortality rates provide, so to say, a 
global quantitative summary of the various
congenital anomalies that appear in the embryonic phase.

\qA{Human infant mortality for all causes of death}

As our starting point we consider infant death
rate curves for humans as shown in Fig.2a,b%
\qfoot{More details about infant mortality
can be found in 
Berrut et al. 2016.}%
.
\qpar

Three striking features of infant mortality rates
appear in Fig.4a,b but before we describe them in detail
we wish to attract the attention of the readers on two
aspects.
(i) Fig.4a shows that the death rates exhibit
little fluctuations.
(ii) Fig.4c shows that the pattern of death rates
remains fairly stable
even when the death rate level changes considerably
as happened between 1923 and 1960. Moreover,
an examination across several countries shows that
these curves remain much the same in all
developed countries. As an illustration, one can
look at the death rate curves for the UK
shown in Berrut et al. (2016)
\qpar

One may think that the first point is 
hardly surprising because the death rate is an average
over a large sample comprising thousands of deaths for each
age interval. 
However, averaging alone cannot explain
the absence of fluctuations as is demonstrated by the
fact that weekly or monthly death rate curves show
fairly large fluctuations. This suggests that
evolution as a function of age is much more stable
than changes in the course of (calendar) time.
As a matter of fact, it will be seen in
Bois et al. (2019a)
that this stability is greater for young-age deaths
than for old-age deaths.
%
%%-----------------------------------------------
%%%%   MORTALITE INFANTILE PR HUMAINS
%%%%   CALCUL FAIT DS: MAPB#MU1960
\begin{figure}[htb]
\centerline{\psfig{width=14cm,figure=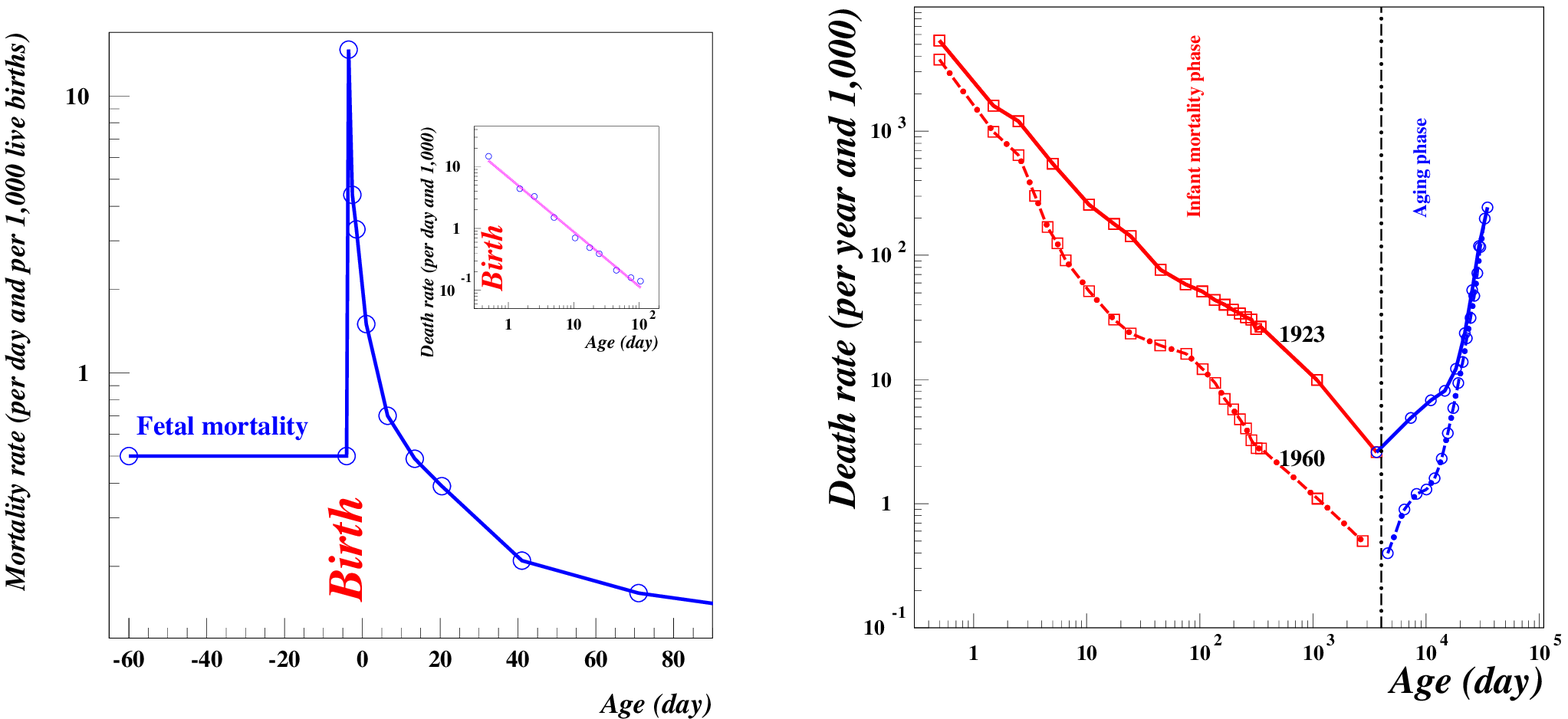}}
\qleg{Fig.\qhu 4a,b\qhv Infant and adult mortality rates
for humans (United States).}
{(a) is for 1923 in the US and the inset is for the same data
in log-log coordinates. Fetal mortality corresponds
to the average level of late fetal mortality (6 to 9 months
pregnancy).
(b) extends until 30,000 days which represents 82 years.
The three main features of infant mortality
are the following: (i) The sharp spike at birth.
(ii) The decrease of infant mortality rate between
birth and the age of 10 followed by subsequent increase.
(iii) The fact that, 
as a function of age $ t $, the decrease
follows an hyperbolic law of the form: $ \mu=A/t^{\gamma} $
with $ \gamma $ of the order of 1.
Note that despite the huge fall of the death rate
between 1923 and 1960 the structure of the two phases
did not change much.  In 1923 $ \gamma=0.65\pm 0.04 $,
whereas in 1960: $ \gamma=1.01\pm 0.08 $ (the error bars
are for a confidence level of 95\%). The change in the slope
from 1923 to 1960
is due to the fact that early mortality is almost time
independent (because mostly due to malformations) 
whereas the mortality at the age of 10 has decreased considerably.
In the interval $ (0,10) $ 
the infant mortality rate is defined as:
$ \mu_b=(1/x_0)\Delta x/\Delta t $, where
$ x_0 $=number of live births, $ \Delta x $=number of deaths
in the age interval $ \Delta t $; this definition is standard for
the interval $ (0,1) $ but  here we extend it to the age interval
$ (0,10) $.
In the expression of
the adult mortality rate $ \mu $, the denominator  
$ x_0 $ is replaced by the number $ x(t) $ of individuals
alive at the beginning of the age interval $ \Delta t $. Actually,
as long as the total infant deaths remain under 10\%, using the
adult definition at all ages
would not make much difference because in this case the infant
age groups are anyway close to $ x_0 $. A last comment is 
in order to say that in the present paper 
the expressions ``death rate'' and ``mortality rate'' are
used as synonyms; sometimes ``death'' is preferred to
``mortality'' just because it is shorter (that is why it is used
in the small inset graph).}
{Sources: 1923 
(a) Under one year: Linder et al. 1947 p.574,
(b) Over one year: Linder et al. 1947, p.150 (gives in fact 1920);
1960 
(a) Under one year: Grove et al. 1968, p.210-211, 
(b) Over one year: Grove et al. 1968, p.318.
}
\end{figure}
%-------------------------------------------------
\qpar

Now we describe the three salient features of the 
shape of the infant death rate curves.
\qee{1} The most impressive feature is certainly the very
sharp spike which coincides with birth. It means 
that the death rate is high immediately
after birth but decreases rapidly in subsequent days and weeks.
\qee{2} In Fig.4a this decrease seems to level off after
the age of 60 days. In fact, the decrease does not
stop but simply becomes slower%
\qfoot{By this expression we mean that a fall from
1,000 to 100 will take place between day 1.5 and 7,
whereas from 10 to 1 it will take from 
day 150 to day 700 (approximately).}%
.
This fall is described by a power law%
\qfoot{Although 
the distinction between power law and
exponential is well known in biology it is not
seen in the same way as in physics.
It is of course obvious that an
exponential falls off faster than
a power law, but one must realize how
massive the difference is.
$$ y_1=1/x,\quad y_2=\exp(-x)\ : x=10\ \rightarrow \ 
y_1=0.1,\ y_2=0.000045 $$
This makes the two functions really different in
nature. For instance, the exponential form
of Gompertz's law {\it absolutely} forbids anybody
to reach the age of 130 years.}
which continues until the age of 3,600 days that is
to say about 10 years. If one considers that the 
maximum life span is about $ T_{\hbox{max}}=100 $ years 
this corresponds to 10\% of  $ T_{\hbox{max}} $.
After the age of 10 years the death rate increases
steadily and exponentially up to $ T_{\hbox{max}} $ 
in accordance with Gompertz's law.
\qpar

Although in medical
language, infant mortality is understood as
the first year after birth, in the present paper
``infant mortality'' refers to the whole phase during
which the death rate decreases. This definition follows
a well established usage in reliability science.
\qee{3} During the infant mortality phase, the human death rate%
\qfoot{Defined as: $ \mu_b=(1/x_0)\Delta x/\Delta t $, where
$ x_0 $=number of life births, $ \Delta x $=number of deaths
in the age interval $ \Delta t $.}
decreases in an hyperbolic way of the form: $ x(t)=A/t^{\gamma} $
where the exponent $ \gamma $ is of the order of 1.

\qA{The age of 10 seen as an equilibrium point between screening and
  wear-out}

If one attributes the downward part of the mortality curve
to a screening process through which individuals with
congenital malformations are eliminated and its upward part
to wear-out, it makes little sense to assume that the first effect
stops at the age of 10 while the second starts at that age.
Certainly the screening continues after 10 and the wear-out starts
immediately after birth. In this perspective, 10 becomes
the equilibrium point between the two effects.

\qA{Infant mortality for specific causes of death}

The graphs of Fig. 4c,d show infant mortality
for specific causes of death, namely viral and bacterial diseases
(of which tuberculosis was the most important instance in the
early 20th century).
Fig 4b, Fig. 4c,d show 
a broad downward trend but in addition for specific
age intervals there are peaks denoting mortality surges.
In fact, these peaks
are also visible on the ``all causes'' curves but only with
poor accuracy because they are overshadowed by the general
trend of all other causes. 

%
%%-----------------------------------------------
%%%%   MORTALITE INFANTILE PR VIRUS ET TB
\begin{figure}[htb]
\centerline{\psfig{width=16cm,figure=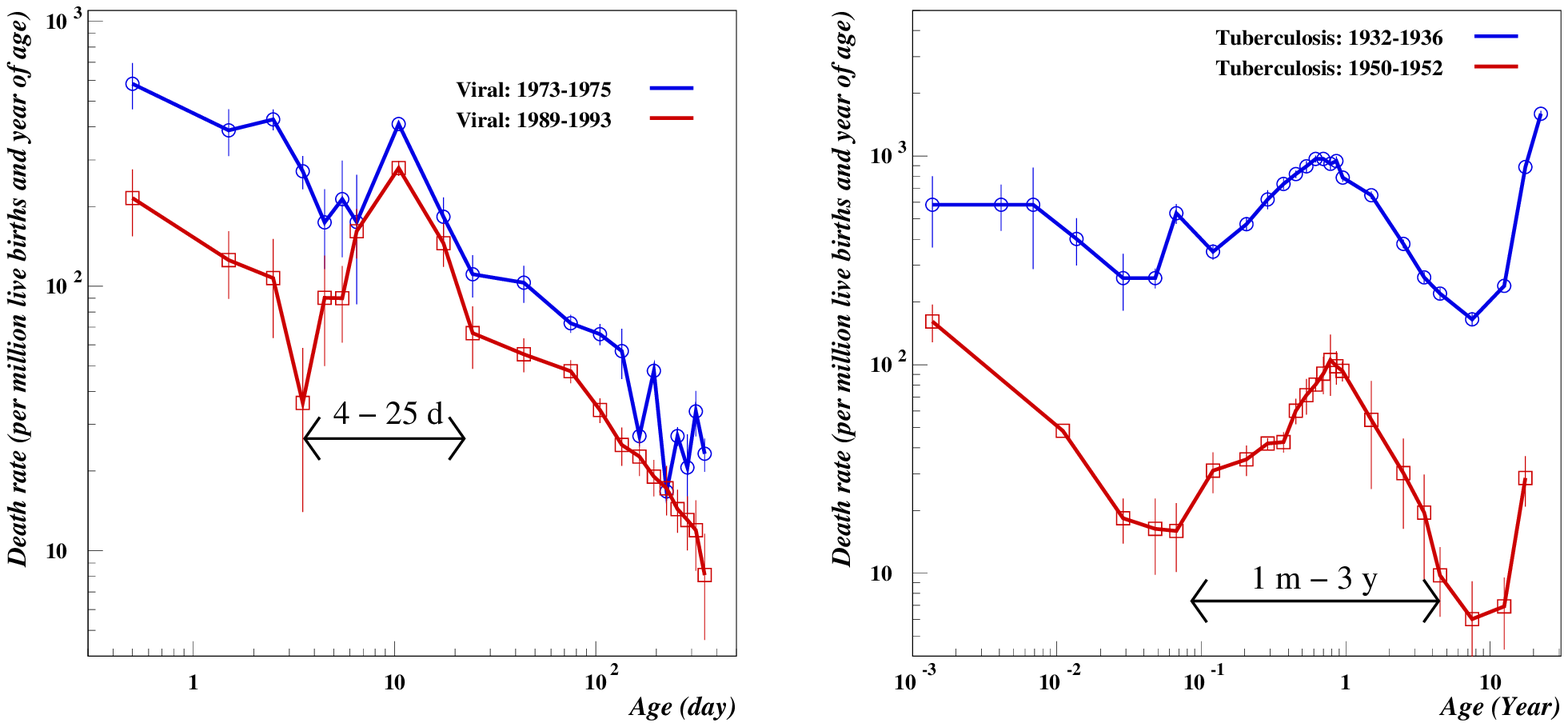}}
\qleg{Fig.\qhu 4c,d\qhv Infant mortality rates
for viral diseases versus tuberculosis.}
{{\bf Left}\quad Although there is a 
general diminution of the death rate
from 1973-1975 to 1989-1993 the peak of the first curve
(in blue)
has an amplitude (ratio of top rate to base rate) of 2 whereas
the second curve (in red) has an amplitude of 5. The
error bars give the standard deviation of the average of 
individual years in the respective age intervals.
{\bf Right}\quad One observes the same phenomenon
as in the graph for viral diseases, namely an
overall diminution coupled with a higher peak in
the more recent time interval: for 1932--1936 the peak
has an amplitude of 3 whereas for 1950-1952 its amplitude
is 6.5. It should be noted that these peaks are also
visible on the total mortality curves but in
attenuated form which means that one needs high accuracy
measurements to detect them.}
{Sources: Vital Statistics for the United States for the
appropriate years; Berrut et al. (2017)}
\end{figure}
%-------------------------------------------------
%

The reason for these peaks is not yet clear but
it is likely that they relate to the
gradual establishment of the immune system. Shortly
after birth the newborn is protected by the antibodies
contained in the breast milk of the mother but this
protection is gradually replaced by the child's own
immune system. Moreover, the immunity provided by the
mother first during pregnancy and then shortly after birth
depends on the diseases that the immune system of the
mother had to face.
\qpar

In other words, these surges
in infant mortality can tell us something about 
special events in infant development that would not
be visible otherwise.

\qI{Conclusion}

\qA{Main results}

The considerable variety of birth defects,
whether lethal or non-lethal, attests that control
mechanisms can be overwhelmed in many ways. However,
the relatively low frequency of each of
these defects (mostly under 1 per
1,000) attests that most of the time the ``manufacturing
process'' works fairly well. 
\qpar

In this paper we have introduced the idea of a third source
of congenital anomalies besides the genetic and environmental
factors. It was called ``manufacturing dispersion'' because 
it consists in the accumulation of small output defects 
in the successive steps of a development process.  
Such a mechanism was shown to be responsible of a substantial
variability even with the two other factors are inactive.
This would solve the mystery of the large proportion
of defects for which no specific source can be identified (as
noted at the beginning of the paper).
\qpar

We have described a number of circumstances 
which are likely to amplify output dispersion:
complex organs, processes which require perfect 
synchronization in time and space, rapid 
and drastic transformations.
\qpar

Clearly one would like to get a better understanding of
the basic mechanisms of manufacturing dispersion.
It is for that purpose
that in a forthcoming paper (Bois et al. 2019b)
we propose two simple physical models which provide
a clearer insight than {\it in vivo} biological organisms..

\qA{Rationale for cross species comparisons}

The dispersion hypothesis led to the prediction
that ``simple'' organisms should have less lethal
congenital anomalies than complex organisms like mammals.
As an illustration consider the following example.
\qpar

In humans, within a few days after birth, 
heart and lungs defects are the main causes of death
(see Fig.5);
lung problems are particularly critical for preterm
newborn. 
%
%%-----------------------------------------------
%%%%  LA PRINCIPALE CAUSE DE MORTALITE INFANTILE 
\begin{figure}[htb]
\centerline{\psfig{width=12cm,figure=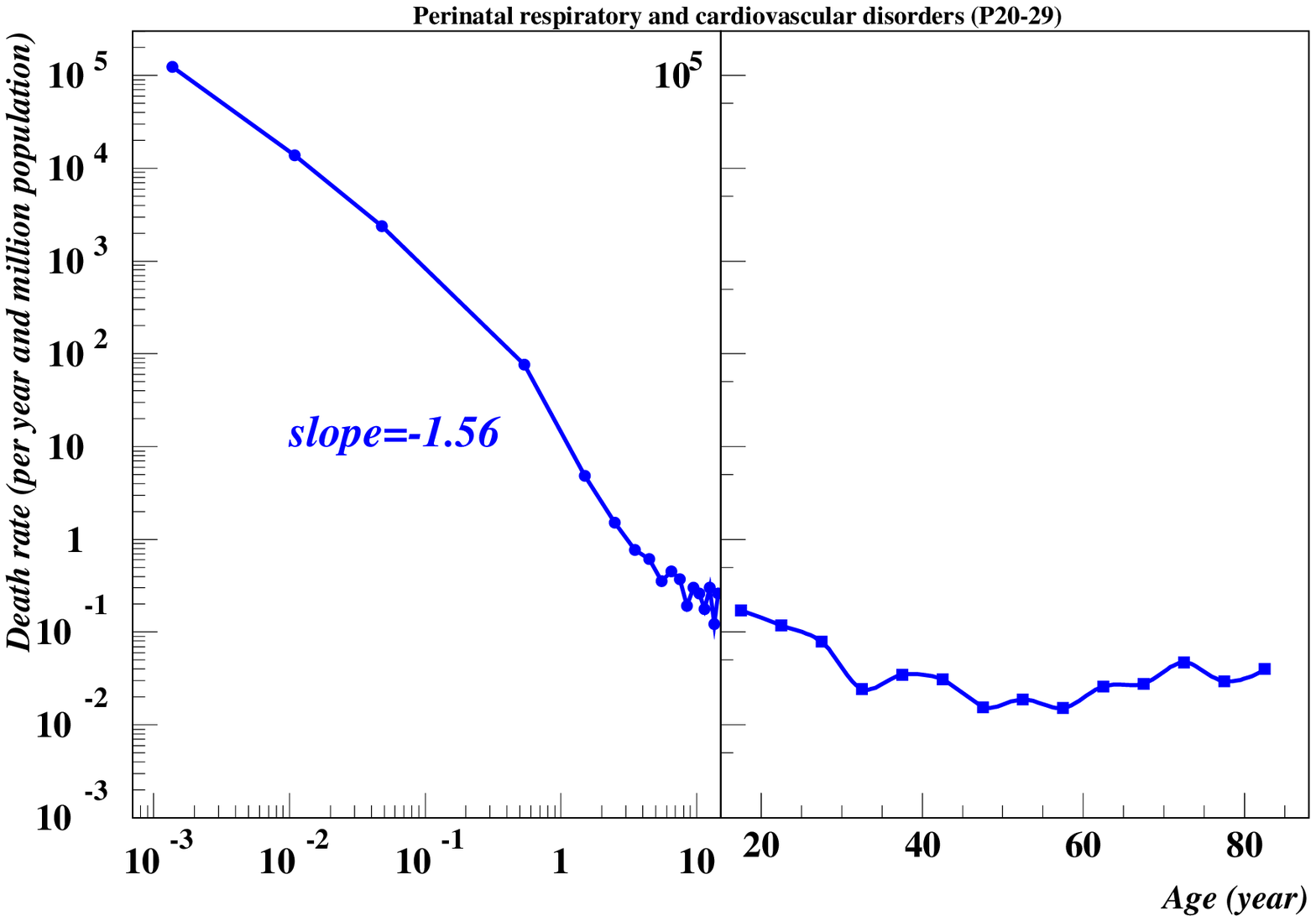}}
\qleg{Fig.\qhu 5\qhv Infant and adult mortality due to
lung and heart congenital abnormalities.}
{There are two noteworthy features. 
(i) The high slope of 1.56 is associated with a high
initial mortality rate (in the first year it is 168 times
higher than for spina bifida)
and reveals 
a drastic screening process. (ii) As a result, this
cause of death is nearly eliminated which explains that the
adult death rate is not increasing with age but instead
fluctuates more or less randomly at a very low level of less
than 100 annual deaths.}
{Source: CDC Wonder: Detailed mortality 1999-2017.}
\end{figure}
%-------------------------------------------------
%
\qpar

In contrast, for rotifers these two causes are
completely non-existent for the simple reason that
rotifers have neither heart nor lungs. Because of their size
(about 0.2mm in length and 0.03mm in diameter) rotifers,
like all other aquatic organisms of similar size or smaller,
receive their oxygen by diffusion through their skin. 
There is of course a similar diffusion process for larger
animals but whereas the concentration jump, $ \Delta c $, 
is the same, the skin thickness, $ \Delta x $, may be
100 times larger, thus giving a diffusion gradient,
$ \Delta c/\Delta x $ some 100 times smaller. Size also
makes blood useless because oxygen can
be brought by diffusion
to all parts of the body. \qL
In short, for rotifers
one does not expect the kind of sharp peak immediately
after birth as observed for humans. Is there nevertheless
an infant mortality phase during which the death rate
decreases? Only observation can tell us. That is why
rotifer mortality will be studied in a companion
paper (Bois et al. 2019a).
\qpar

Incidentally, it can be observed that the diffusion mechanism
works not only for microscopic organisms but also
for centimeter-size organisms on the condition that they
are formed of thin layers.
That is the case for: (i) sponges consisting
of a single cell layer or (ii) jelly fish whose body is a layer
not more than a few cells thick. In all these organisms
gases, nutrients, and wastes are exchanged by diffusion.
Thus, as a conjecture, one would expect their infant mortality
curve to start similarly as the one of rotifers.
\qpar

More broadly, 
it is in order to test such predictions that
we started a research program consisting in the measurement
of infant mortality across species.

\appendix

\qI{Appendix A. Estimating the strength of genetic factors}

It is probably not far from the truth to say that 
nowadays some 90\% of the research papers in biology 
are to some extent focused on genetics. This is surprising
because, as explained in a review paper published
in the ``New York Times'' (Kolata 2006),
demographic and epidemiological research shows 
that for most human characteristics (e.g. lifespan or
diseases) there is only a very loose genetic influence. 
\qpar

Here we are interested in birth defects.
Because they are not affected by all life incidents
(which differ from person to person) one may think that
there is a firmer ground for genetic influence.
Currently, it seems to be a well accepted axiom that
most malformations have a genetic origin. At least this is 
the implication of papers like the study by Ahmed et al. (2017)
which, for all separate variants of finger malformations,
lists the genes which seem responsible.
Under such an assumption, monozygotic twins should
have the same malformations. We will see below that this
is far from true.
\qpar

Before focusing on the twin methodology let us briefly
examine some aspects of harmful mutations leading to
anomalies.

\qA{Mutations and repair mechanisms}

In a living organism harmful mutations can occur at three levels.
(i) Germ cells.
(ii) Stem cells, i.e. cells no yet
differentiated into specific organ types.
(iii) Fully functional differentiated cells existing in various
organs. 
The last two types are called somatic mutations for they
are not passed on to children.
\qpar

In a long term perspective the most serious cause of concern
are of course the germ cell mutations because, unless there
is a repair mechanism, they will be passed over from generation
to generation and will accumulate%
\qfoot{There may be many external mutation factors but one 
that has existed without any doubt since the beginning
of life on Earth consists in high energy cosmic rays.}%
. 
So, the existence of
effective repair mechanisms has been a natural assumption
among biologists long before it was eventually
demonstrated in a work honored by a Nobel award in 2015.
If there are repair mechanisms it means that the static picture 
with a rigid connection between defective genes 
and abnormalities must be replaced by a dynamic 
vision.

\qA{The twin methodology for assessing the strength of genetic
factors}

A methodology based on twin data which permits to ascertain
the role of genetic factors in the occurrence of malformations
(or more generally of any disease or trait) has been
developed by several authors, e.g. Hrubec et al. (1981)
and Tishler et al. (2007).  However, as the method is used
differently in each specific application, we summarize
in this appendix the variables and reasoning which are
most convenient for our purpose.
\qpar

Before giving a formalized presentation for a large
sample of twin pairs it may be useful to describe a 
specific case consisting in the occurrence of breast
cancer in monozygotic twins. A team of  Czech researchers
followed 5 monozygotic pairs of twins over a long time
period of up to two decades. They made the following observations
(Hlad\`{\i}kov\'a et al. 2013).
\qbu Pair 1=(breast cancer at age 54 versus ovarian cancer at age 43)
\qbu Pairs 2,3,4,5=(breast cancer at a median age of 44 versus no cancer)
\qpar

The authors conclude that ``environmental factors play an important 
role in breast cancer development''. Instead of
mysterious ``environmental factors'' such outcomes can also 
result from 
a random dispersion of manufacturing outputs.
\qpar
Next, we consider this problem in a more general way.
\qpar

The starting point is a dataset for a sample comprising
$ M $  monozygotic (MZ) twins and $ D $ dizygotic (DZ) twins.
Secondly, one focuses on the frequency of a specific 
congenital malformation.
This leads to define and compute the following variables.
\qbu  Concordant pairs, i.e. pairs in which both twins have the 
malformation;
we denote their number by $ c_m $ and $ c_d $ respectively
for MZ and DZ twins.
\qbu ``Discordant'' pairs, i.e. pairs in which one child has the 
malformation 
but not the other; we denote their number by $ d_m $ and $ d_d $
respectively for MZ and DZ twins.
\qpar

In addition, we denote the probability of the 
malformation in the general  population by $ p $. A typical
order of magnitude for $ p $ is 1 per 1,000 that is to say: 
$ p=10^{-3} $.
\qpar

Ideally, for a malformation that is 100\% genetically determined,
among MZ twins
there should be no discordant pairs, i.e. $ d_m=0 $. Thus,
if we introduce the ratio $ g_m=c_m/(c_m+d_m) $ it will be equal
to 1. 
\qpar

In contrast, for DZ twins there may be some discordant pairs,
i.e. $ d_d>0 $. Thus, for $ g_d=c_d/(c_d+d_d) $ one gets:
$ g_d < 1 $, in other words: $ g_d<g_m $;
this last inequality is also expected to hold at least approximately 
for malformations in which
genetic determination is less than 100\%.
\qpar

For a malformation which has no genetic basis at all,
the probability for both twins to have it would be $ p^2 $,
whereas the probability for only one having it
would be: $ p(1-p) $; as usually $ p $ is of the order
of one per thousand  the factor $ 1-p $ can be approximated by 1.\qL
Thus, 
$$ c_m=M p^2,\ d_m=M p \ \rightarrow \ 
g_m \simeq p^2/(p^2+p)=p/(p+1)\simeq p $$
\qpar

Naturally, in this case the expectations for DZ twins are 
the same as for MZ twins.
\qpar

In short, the strength of genetic factors can be estimated in two
ways: 
\qee{i} How close is $ g_m $ to 1? 
It turns out that for most congenital malformations $ g_m $ is
smaller than $ 0.3 $. In the previous cancer example, $ c_m=0 $
because even for pair 1 there are different cancers%
\qfoot{If one is only interested in whether there is cancer or not
then $ c_m=1 $ and $ g_m=1/5=0.2 $.}%
,
thus $ g_m=0 $.
\qee{ii} How much is $ g_m $ larger than $ g_d $?
This can be expressed by the ratio: $ g'=g_m/g_d $.
In the cancer example: $ g'=0 $.
\qpar

These conclusions are summarized in Table A1a.

%
%%-----------------------------------------------
\begin{table}[htb]

\small
\centerline{\bf Table A1a: Twin variables for estimating the
strength of genetic factors in malformation occurrences.}

\vskip 5mm
\hrule
\vskip 0.7mm
\hrule
\vskip 0.5mm
$$ \matrix{
\hbox{}\hfill & \hbox{MZ} & \hbox{MZ} & \hbox{MZ} & 
\quad & \hbox{DZ} & \hbox{DZ} & \hbox{DZ} & \quad & 
\hbox{MZ/DZ} \cr
\hbox{}\hfill & \hbox{Concord.} & \hbox{Discord.} & 
\hbox{Ratio} & 
\quad & \hbox{Concord.} & \hbox{Discord} & \hbox{Ratio} & \quad & 
\hbox{} \cr
\hbox{}\hfill & \hbox{pairs} & \hbox{pairs} & 
\hbox{} & 
\quad & \hbox{pairs} & \hbox{pairs} & \hbox{} & \quad & 
\hbox{} \cr
\qtb
\hbox{}\hfill & c_m & d_m & 
g_m & 
\quad & c_d & d_d & g_d & \quad & 
g'=g_m/g_d \cr
\noalign{\hrule}
\qth
\hbox{100\% genetic}\hfill & c_m & d_m=0 & 
g_m=1 & 
\quad & c_d & d_d>0 & g_d<g_m & \quad & 
g'>1 \cr
\qtb
\hbox{\phantom{10}0\% genetic}\hfill & Np^2 & Np & 
g_m=p & 
\quad & Np^2 & Np & g_d=p & \quad & 
g'\simeq 1 \cr
\noalign{\hrule}
} $$
\vskip 1.5mm
Notes: $ N $ is the population of the sample.
MZ means monozygotic (true twins) and it corresponds to the 
index $ m $.
DZ means dizygotic and it corresponds to the index $ d $.
``Concord.'' means ``Concordant'' (corresponds to the variable $ c $). 
``Discord.'' means ``Discordant'' (corresponds to the variable $ d $).
As an example of the notations, 
the variable $ c_d $ represents
``concordant pairs of dizygotic twins.
$ g_m $ and $ g_d $ have the following definitions:
$ g_m=c_m/(c_m+d_m),\ g_d=c_d/(c_d+d_d) $.
$ p $ is the probability of the malformation in the general
population; it is assumed that $ p\ll 1 $ (usually $ p\simeq 10^{-3}
$).
High strength of genetic factors is associated with $ g_m $ close
to 1 and $ g' $ higher than 1, whereas low strength is associated
with $ g_m $ much smaller than 1 and $ g' $ close to 1.  
%\qL
%{\it Sources:}
\vskip 2mm
\hrule
\vskip 0.7mm
\hrule
\end{table}
%%-----------------------------------------------

Inserting the values of $ c_m,d_m,c_d,d_d $ given in Yu et al (2019,
Table 2) one gets the results shown in Table A1b.

%
%%-----------------------------------------------
\begin{table}[htb]

\small
\centerline{\bf Table A1b: Estimates of the
strength of genetic factors in malformation occurrences.}

\vskip 5mm
\hrule
\vskip 0.7mm
\hrule
\vskip 0.5mm
$$ \matrix{
\hbox{Birth}\hfill & p & \hbox{MZ} & \hbox{MZ} & \hbox{MZ} & 
  \hbox{DZ} & \hbox{DZ} & \hbox{DZ} &  
\hbox{MZ/DZ} \cr
\hbox{defect}\hfill & \hbox{per} & \hbox{Concord.} & \hbox{Discord.} & 
\hbox{Ratio} & 
  \hbox{Concord.} & \hbox{Discord} & \hbox{Ratio} &  
\hbox{} \cr
\hbox{}\hfill & 1,000 &\hbox{pairs} & \hbox{pairs} & 
\hbox{} & 
 \hbox{pairs} & \hbox{pairs} & \hbox{} & 
\hbox{} \cr
\hbox{}\hfill && c_m & d_m & 
g_m= & 
 c_d & d_d & g_d=&  
g'=g_m/g_d \cr
\vrule height 0pt depth 20pt width 0pt
\hbox{}\hfill &&  &  & 
\displaystyle {c_m\over c_m+d_m} & 
  &  & \displaystyle {c_d\over c_d+d_d}& 
g'=g_m/g_d \cr
\noalign{\hrule}
\qth
\hbox{Oral cleft}\hfill & 2 & 2 & 7 & 
22\% & 
 2 & 36 & 5.3\% &  
4.1 \cr
\qtb
\hbox{Spina bifida}\hfill & 2 & 1 & 16 & 
5.9\% & 
 0 & 33 & 0\% &  
- \cr
\hbox{}\hfill &  &  &  & 
 & 
  &  &  &  
 \cr
\hbox{Club foot}\hfill & 4 & 5 & 17 & 
 22\% & 
 4 & 82 & 4.6\%  &  
 4.8 \cr
\hbox{Strabism}\hfill & 18 & 33 & 161 & 
 17\% & 
 27 & 412 & 6.5\%  &  
 2.6 \cr
\hbox{}\hfill &  &  &  & &   &  &   &   \cr
\qtb
\hbox{\color{blue}Average}\hfill & \color{blue}6.5 &  &  & 
 \color{blue}16.7\%& 
  &  & \color{blue}4.10\%  &  
 \color{blue} 3.83 \cr
\noalign{\hrule}
} $$
\vskip 1.5mm
Notes: Although for oral cleft and
spina bifida the numbers of cases are somewhat too small
the fact that among MZ pairs
there are much more discordant pairs than
concordant pairs (which translates in a value of $ g_m $
much lower than 1)
shows a loose genetic determination. The results for $ g' $
are only significant for strabismus; for the other
defects there are too few DZ cases. 
\qL
{\it Sources: The data are for 6,752 monozygotic twin pairs
and 13,310 dizygotic twin pairs from the California twin program
covering 1957--1982 (Yu et al. 2019)}.
\vskip 2mm
\hrule
\vskip 0.7mm
\hrule
\end{table}
%%-----------------------------------------------

The estimates show that for all malformations 
the strength of genetic
factors is far from 100\%; in other words there is room
for other factors than heredity particularly for
environmental factors and output dispersion.
According to the $ g_m $ criterion, the strength 
of genetic factors rank as follows (from high to low):
oral cleft, club foot, strabismus, spina bifida;
according to the $ g' $ criterion the ranking is:
club foot, oral cleft, strabismus (not defined for spina
bifida).
\qpar

In Table A1b we see that: (i) $ g_m> p $,
(ii) $ g_d<g_m $ and (iii) $ g'> 1 $  which suggests that 
genetic factors play a role in the malformations.
However, the fact that on average for the 4 malformations
$ g_m=0.16\pm 0.04 $ which is well below 1
show that genetic determination is rather weak.
In other words, other factors may be at work.

\qA{Strength of genetic factors in cancer}

So far, we have examined birth defects. Although
it is at birth that these defects become visible, in fact 
they appear earlier during pregnancy.
On the contrary, cancer appears late in the course of life.
Therefore, one can expect important contributions
of somatic mutations and environmental factors.
It is for the purpose of comparison that we study this case.

%
%%-----------------------------------------------
\begin{table}[htb]

\small
\centerline{\bf Table A1c: Estimates of the
strength of genetic factors in cancer.}

\vskip 5mm
\hrule
\vskip 0.7mm
\hrule
\vskip 0.5mm
$$ \matrix{
\hbox{}\hfill &  \hbox{MZ} & \hbox{MZ} & \hbox{MZ} & 
\quad & \hbox{DZ} & \hbox{DZ} & \hbox{DZ} & \quad & 
\hbox{MZ/DZ} \cr
\hbox{Type of}\hfill  & \hbox{Concord.} & \hbox{Discord.} & 
\hbox{Ratio} & 
\quad & \hbox{Concord.} & \hbox{Discord} & \hbox{Ratio} & \quad & 
\hbox{} \cr
\hbox{cancer}\hfill  &\hbox{pairs} & \hbox{pairs} & 
\hbox{} & 
\quad & \hbox{pairs} & \hbox{pairs} & \hbox{} & \quad & 
\hbox{} \cr
\qtb
\hbox{}\hfill & c_m & d_m & 
g_m \ (\%) & 
\quad & c_d & d_d & g_d \ (\%)& \quad & 
g'=g_m/g_d \cr
\noalign{\hrule}
\qth
\hbox{\bf Specific cancers}\hfill  &  &  & & \quad &  &  &  & \quad &  \cr
\hbox{Lung}\hfill  & 1 & 49 & 
2.0\% & 
\quad & 3 & 112 & 2.6\% & \quad & 
0.77 \cr
\hbox{Stomach}\hfill  & 2 & 74 & 
4.9\% & 
\quad & 4 & 138 & 2.8\% & \quad & 
0.93 \cr
\hbox{Colon}\hfill & 8 & 153 &   
 2.6\% & 
\quad & 13  & 191 & 4.3\% & \quad & 
 1.16\cr
\hbox{Breast}\hfill  & 22 & 257 & 
 7.9\% & 
\quad & 23 & 467 & 4.7\%  & \quad & 
 1.68 \cr
\hbox{Cervix}\hfill  & 30 & 242 & 
 11.0\% & 
\quad & 27 & 412 & 5.1\%  & \quad & 
 2.16 \cr
\hbox{Prostate}\hfill  & 19 & 137 & 
 12.2\% & 
\quad &7  &299  & 2.3\%  & \quad & 
 5.30 \cr
\hbox{\color{blue}Average}\hfill  &  &  & 
 \color{blue} 6.8\% & 
\quad &  &  & \color{blue}3.6\%  & \quad & 
 \color{blue}2.00 \cr
\hbox{}\hfill  &  &  & 
 & 
\quad &  &  &   & \quad &  \cr
\qtb
\hbox{\bf All cancers}\hfill  & 182 & 1306 & 
 12.2\% & 
\quad & 257 & 2351 & 3.6\%  & \quad & 
 1.23 \cr
\noalign{\hrule}
} $$
\vskip 1.5mm
Notes: 
The variables $ c_m,c_d,d_m,d_d,g' $ are defined in the text.
``Concord'' means ``Concordant'' (i.e. same disease in
each twin of a pair);
``Discord'' means ``Discordant''. In the ``Specific cancers''
cases ``concordant'' means the same specific kind of cancer whereas
in the ``All cancers'' row ``concordant'' means ``any kind of cancer''.
The cancers are ranked by order of increasing values
of $ g' $, that is to say increasing strength of genetic factors.
The ``All cancers'' row includes more cases than the 6 types
listed in the table.
\qL
{\it Sources: The data are for 23,386 twin pairs from
the ``Swedish Twin Registry'' covering the years
1959--1961 and 1970--1972 (Ahlbom et al. 1997)}.
\vskip 2mm
\hrule
\vskip 0.7mm
\hrule
\end{table}
%%-----------------------------------------------
%

Table A1c gives estimates for the strength of genetic factors
in cancer. Whether or not cancer can be seen as resulting from
a congenital defect of the immune system is a matter
of perspective. On average the estimates show that the 
genetic component is weaker than for the malformations
given in Table A1b.
\qpar

When the concordance of monozygotic and dizygotic twin pairs are
approximately of same value, i.e. $ g'\sim 1 $, it suggests
a small influence of  genetic factors. In such a situation 
one must check if this common value is higher than what would be
expected on a purely random basis. 
\qpar

For all cancers except cervix and prostate cancer,
on account of $ g_m\simeq g_d $ 
there is little genetic influence. As the prevalence for all cancers
is about $ p=6\% $ in the population over 15, Table A1c shows that
$ g_m=12.2\% $ is (slightly) higher than the random threshold of $ 6\% $.
This suggests that family similarities may play a role,
e.g. obesity, stress due to living or working conditions and so on. 
\vskip 4mm

{\bf Acknowledgments} \quad The authors express their gratitude
to Drs. Kun Wang and Luc Westphal for their help and interest.

\vskip 5mm

{\bf References}

\qparr
Ahlbom (A.), Lichtenstein (P.), Malmstr\"om (H.), Feychting (F.),
Hemminki (K.), Pedersen (N.L.) 1997:
Cancer in twins: genetic and nongenetic familial
risk factors.
Journal of the National Cancer Institute 89,4,287-293.

\qparr
Alix (M.) 2016: Etude de la variabilit\'e de l'embryog\'en\`ese
chez la perche commune: d\'eveloppement d'approches alternatives.
[Embryogenesis of the European perch (Perca fluviatilis):
alternative approaches.]
PhD Thesis presented at the University of Lorraine on 15 December
2008.

\qparr
Berrut (S.), Pouillard (V.), Richmond (P.), Roehner (B.M.) 2016:
Deciphering infant mortality.
Physica A 463,400-426.

\qparr
Berrut (S.), Richmond (P.), Roehner (B.M.) 2017:
Age spectrometry of infant death rates as a probe of
immunity: Identification of two peaks due to viral and
bacterial diseases respectively.
Physica A 486,915-924.

\qparr
Biobaku (K.T.), ADELEYE (O.E.) 2010:
Two hens mutually brooding: a rare behaviour in 
{\it Gallus domesticus}.
Science World Journal 5,3,21-22.

\qparr
Boas (H.M.) 1918. Inheritance of eye color in man. 
American Journal of Physical Anthropology 2,15-20.

\qparr
Bois (A.),
Garcia-Roger (E.M.),
Hong (E.),
Hutzler (S.),
Ali Irannezhad (A.),
Mannioui (A.),
Richmond (P.),
Roehner (B.M.),
Tronche (S.) 2019a: 
Infant mortality across species.
A global probe of congenital abnormalities.
Preprint April 2019.

\qparr
Bois (A.),
Garcia-Roger (E.M.),
Hong (E.),
Hutzler (S.),
Ali Irannezhad (A.),
Mannioui (A.),
Richmond (P.),
Roehner (B.M.),
Tronche (S.) 2019b: 
Physical models of infant mortality.
Implications for biological systems.
Preprint June 2019.

\qparr
Burdett (I.D.J.), Kirkwood (T.B.L.), Whalley (J.B.) 1986:
Growth kinetics of individual {\it Bacillus subtilis} cells
and correlation with nucleoid extension.
Journal of Bacteriology 167,1,219-230.

\qparr
Chen (J.),
Huang (X.), 
Wang (B.), 
Zhang (Y.),
Rongkavilit (C.),
Zeng (D.),
Jiang (Y.),
Wei (B.), 
Sanjay (C.),
McGrath (E.) 2018:
Epidemiology of birth defects based on surveillance data from
2011-2015 in Guangxi, China: comparison across five major ethnic
groups.
BMC [Bio-Med Central] Public Health 18,1008.

\qparr
Crymes (W.B.), Zhang (D.), Ely (B.) 1999: 
Regulation of podJ expression during the {\it Caulobacter crescentus} 
cell cycle.
Journal of Bacteriology 181,13,3967-3973. 

\qparr
Danielsen (R.), Aspelund (T.), Harris (T.B.), Gudnason (V.) 2014:
The prevalence of aortic stenosis in the elderly in Iceland and
predictions for the coming decades: The AGES-Reykjavík study.
International Journal of Cardiology 176,3,916-922.

\qparr
Depree (J.A.), Geoffroy (P.S.) 2001:
Physical and flavor stability of mayonnaise.
Trends in Food Science and Technology 12,5,157-163.

\qparr
Drummond (D.A.), Wilke (C.O.) 2009:
The evolutionary consequences of erroneous protein synthesis.
Nature Reviews Genetics 10,10,715-724.

\qparr
Dubrova (E.) 2013: Fault-tolerant design.
Springer, Berlin.

\qparr
Fairchild (B.D.), Christensen (V.L.), Grimes (J.L.),
Wineland (M.J.), Bagley (L.G.) 2002: Hen age relationship with
embryonic mortality and fertility in commercial turkeys.
The Journal of Applied Poultry Research 11,3,260-265.

\qparr
Feldkamp (M.L.), Carey (J.C.), Byrne (J.L.B.), Sergey Krikov (S.)  
Botto (L.D.) 2017:
Etiology and clinical presentation of birth defects: 
population based study.
British Medical Journal 357,j2249.

\qparr
French (F.E.), Bierman (J.M.) 1962: Probabilities of fetal 
mortality. Public Health Reports 77,10,835-847.

\qparr % 2 morts ds les premieres 12h, n=72
Garcia-Roger (E.M.), Mart\'{i}nez (A.), Serra (M.) 2006:
Starvation tolerance of rotifers produced from
parthenogenetic eggs and from diapausing eggs: a life table
approach.
Journal of Plankton Research 28,3,257-265.

\qparr
Gerhard (G.S.), Kauffman (E.J.), Wang (X.), Stewart (R.), 
Moore (J.L.), Kasales (C.), Demidenko (E.), Cheng (K.C.) 2002: 
Life spans and senescent phenotypes
on two strains of Zebrafish (Danio rerio). 
Experimental Gerontology 37,1055-1068.

\qparr % life span, n=10
Gopakumar (G.), Jayaprakas (V.) 2004: Life table parameters
of {\it Brachionus plicatilis} and {\it B. rotundiformis}
in relation to salinity and temperature.
Journal of the Marine Biological Association of India 46,1,21-31.

\qparr
Greenough (R.B.) 1925: Varying degrees of malignancy in cancer of the breast. 
Journal of Cancer Research, 1925,453-463.

\qparr
Grove (R.D.), Hetzel (A.M.) 1968:  Vital statistics rates in the United
States, 1940–1960. 
United States Printing Office, Washington, DC.

\qparr
Gunton (K.B.), Wasserman (B.N.), DeBenedictis (C.) 2015: 
Strabismus. Primary care 42,3,393-407.

\qparr
Hirschmann (W.B.) 1964: Profit from the learning curve.
Harvard Business Review, 42,1,125-139.

\qparr
Hlad\`{\i}kov\'a (A.), Plevov\'a (P.), Mach\'a\u{c}kov\'a (E.) 2013:
Breast cancer in monozygotic twins [in Czech].
Klin Onkol. 26,3,213-217.

\qparr
Hrubec (Z.), Neel (J.V.) 1981:
Familial factors in early deaths: twins followed 30 years to ages
51-61 in 1978.
Human Genetics 59,1,39-46.

\qparr
Hutt (F.B.) 1929: Studies in embryonic mortality in the
fowl. I. The frequency of various malpositions of the chick embryo
and their significance. Proceedings of the Royal Society
of Edinburgh 49,II,118-130. 

\qparr
ISA (Institut de S\'election Animale) 2009: From egg to chicken.
Hatchery manual. Boxmeer (Netherlands). [A 46-pages handbook
published by ISA which is a subsidiary of the 
``Hendrix Genetics Company''.]

\qparr
Jennings (H.S.) 1916: Heredity, variation and the results
of selection in the uniparental reproduction of 
{\it Difflugia corona}. Genetics 1,407-534.

\qparr % plat jusqu'a 356h, n=240
Jian (X.), Tang (X.), Xu (N.), Sha (J.) YouWang (Y.) 2017:
Responses of the rotifer Brachionus plicatilis to flame retardant
(BDE-47) stress.
Marine Pollution Bulletin 116,1-2,298-306.

\qparr
Joe (B.B.) 2004: Growth response of {\it Euglena gracilis}
and {\it Selenastrum capricornutum} in response to pH.
Semantic Scholar (22 March 2004).

\qparr
Johannsen (W.) 1903: Ueber Erblichkeit in Populationen und
reinen Linien. Ein Beitrag zur Beleuchtung schwebender
Selektionsfragen. [About heredity in populations and
pure lines. A contribution to the understanding of
pending questions about selection.]
Gustav Fisher, Jena.

\qparr % plat jusqu'a 3j quel que soit la temp, n=120
Johnston (R.K.), Snell (T.W.) 2016: 
Moderately lower temperatures greatly extend the lifespan of Brachionus
manjavacas (Rotifera): Thermodynamics or gene regulation?
Experimental Gerontology 78,12-22.

\qparr
Joshi (P.S.) 1988a: Influence of salinity on population
growth of a rotifer, {\it Brachionus plicatilis}.
Journal of the Indian Fisheries Association 18,75-81.

\qparr
Joshi (P.S.) 1988b: Mass culture of {\it Brachionus plicatilis}.
Master of Science Dissertation, University of Bombay (84 p.)
[available on line]

\qparr
Kioumourtzoglou (M.-A.), Coull (B.A.), O'Reilly (E.J.), Ascherio (A.), 
Weisskopf (M.G.) 2018: Association of exposure to 
diethylstilbestrol [DES]
during pregnancy with multigenerational neurodevelopmental deficits.
Journal of the American Medical Association (JAMA), Pediatrics. 
172,7,670-677.

\qparr
Kobitski (A.Y.), Otte (J.C.), Takamiya (M.), 
Sch\"afer (B.), Mertes (J.), Stegmaier (J.), 
Rastegar (S.), Rindone (F.), Hartmann (V.), 
Stotzka (R.), Garc\'{\i}a (A.), Wezel (J. van), Mikut (R.), 
Str\"ahle (U.), Nienhaus (G.U.) 2015: 
An ensemble-averaged, cell density-based digital model of zebrafish
embryo development derived from light-sheet microscopy data with
single-cell resolution.
Nature Scientific Reports 5,8601.

\qparr
Knight (K.) 2016: Zebrafish larvae learn to hunt using lateral line 
in the dark. 
Journal of Experimental Biology 219,465-466.

\qparr
Kohler (I.V.), Preston (S.H.), Lackey (L.B.) 2006:
Comparative mortality levels among selected species of captive
animals.
Demographic Research 15,14,413-434.

\qparr
Kolata (G.) 2006: Live long? Die young? Answer isn't just in genes.
New York Times 31 August 2006.

\qparr
Lashley (K.S.) 1915: Inheritance in the asexual reproduction of hydra.
The Journal of Experimental Zoology 19,157-210.

\qparr
Laub (M.T.), McAdams (H.H.), Feldblyum (T.), Fraser (C.M.), 
Shapiro (L.) 2000: Global analysis of the genetic network
controlling a bacterial cell cycle. 
Science 290,2144-2154.

\qparr
Linder (F.E.), Grove (R.D.) 1947: Vital statistics rates in
the United States,1900-1940.
United States Printing Office, Washington, DC, 1947.

\qparr
Middleton (A.R.) 1915a: Heritable variations and the results
of selection in the fission rate of {\it Stylonychia pustulata.}
Journal of Experimental Zoology 19,451-503.

\qparr
Middleton (A.R.) 1915b: Heritable variations and the results
of selection in the fission rate of {\it Stylonychia pustulata.}
Proceedings of the National Academy of Sciences, 4 November 1915.\qL
[This paper is a summary of the previous one.]

\qparr
Noyes (B.) 1923: Experimental studies on the life history
of a rotifer reproducing parthenogenetically ({\it 
Proales decipiens}).
Journal of Experimental Zoology 35,2,225-255.\qL
[It is a thesis submitted to John Hopkins University,
Baltimore.]

\qparr
Papoulis (A.) 1965: Probability, random variables, and
stochastic processes. 
McGraw-Hill, Tokyo.

\qparr
Patey (D.H.), Scarff (R.W.) 1928: The position of histology in the
prognosis of carcinoma of the breast.
The Lancet 211,5460,801-804.

\qparr
Pearl (R.), Miner (J.R.) 1935: Experimental studies on the duration
of life. XIV. The comparative mortality of certain lower organisms.
The Quarterly Review of biology 10,1,60-79.\qL
[The paper contains data for Hydra fusca, 

\qparr
Pearl (R.) Park (T.), Miner (J.R.) 1941: Experimental studies on the
duration of life. XVI Life tables for the flour beetle 
{\it Triboleum confusum Duval}.
The American Naturalist 75,756,5-19. 

\qparr
Pe\~nuela (A.), Hernandez (A.) 2018: 
Characterization of embryonic mortality in broilers.
Revista MVZ (Medecina, Vetenaria,Zootecnia) C\'ordoba 23,1,6500-6513.

\qparr
Pouillard (V.) 2015: En captivit\'e. 
Vies animales et politiques humaines dans les jardins zoologiques 
du XIXe si\`ecle \`a nos jours : 
m\'enagerie du Jardin des Plantes, zoos de Londres et Anvers.
[In captivity. Zoo management and animal lives in the
zoological gardens of Paris, London and Antwerp
from the 19th century to 2014.]. PhD thesis.
Universit\'e libre de Bruxelles and University of Lyon 3.

\qparr
Romanoff (A.) 1949: Critical periods and causes of death in
avian embryonic development.
The Auk 66,3,264-270.

\qparr
Rombough (P.J.) 1998: 
Partitioning of oxygen uptake between the gills and skin in fish
larvae: a novel method for estimating cutaneous oxygen uptake.
Journal of Experimental Biology 201,11,1763-1769.

\qparr
Rideout (R.M.), Trippel (E.A.), Litvak (M.K.) 2004:
Predicting haddock embryo viability based on early cleavage patterns.
Aquaculture 230,215-228.

\qparr
Sahin (T.) 2001: Larval rearing of the Black Sea Turbot, 
{\it Scophthalmus maximus} (Linnaeus, 1758), under
laboratory conditions.
Turkish Journal of Zoology 25,447-452.

\qparr
Schaefer (B.M.), Lewin (M.B.), Stout (K.K.), 
Gill (E.), Prueitt (A.), Byers (P.H.), Otto (C.M.) 2007:
The bicuspid aortic valve: an integrated phenotypic classification of
leaflet morphology and aortic root shape.
British Medical Journal Heart 94,12.

\qparr
Shylakhovenko (V.A.),  Olishevsky (S.V.), Kozak (V.V.), 
Yanish (Y.V.), Rybalko (S.L.) 2003:
Anticancer and immunostimulatory effects of nucleoprotein fraction of 
{\it Bacillus subtilis}. 
Experimental Oncology. 25,119-123.

\qparr
Stocking (R.J.) 1915a: Variation and inheritance of abnormalities
occurring after conjugation in {\it Paramecium caudatum.}
Journal of Experimental Zoology 19,387-449.

\qparr
Stocking (R.J.) 1915b: Variation and inheritance in abnormalities
occurring after conjugation in {\it Paramecium caudatum.}
Proceedings of the National Academy of Sciences, 4 November 1915.\qL
[This paper is a summary of the previous one.]

\qparr % plat jusqu'a 50h, nb pas donne
Sun (Y.), Hou (X.), Xue (X.), Zhang (L.), Zhu (X.), 
Huang (Y.), Chen (Y.), Yang (Z.) 2017:
Trade-off between reproduction and lifespan of the rotifer Brachionus
plicatilis under different food conditions.
Scientific Reports 7,1.

\qparr
Tishler (P.V.), Carey (V.J.) 2007:
Can comparison of MZ- and DZ-twin concordance rates be used invariably
to estimate heritability?
Twin Research and Human Genetics 10,5,712-717.

\qparr
Tomasetti (C.), Vogelstein (B.) 2015: Variation in cancer risk
among tissues can be explained by the number of stem cell divisions.
Science 347,78-81.

\qparr
Uchida (Y.), Uesaka (M.), Yamamoto (T.), Takeda (H.), Irie (N.)
2018: Embryonic lethality is not sufficient to explain hourglass-like
conservation of vertebrate embryos.
EvoDevo 9,7,1-11.

\qparr
Vyas (R.), Verma (S.), Malu (V.K.) 2016: 
Distribution of congenital malformations at birth in a tertiary care
hospital in North-Western Rajasthan.
International Journal of Reproduction, Contraception, Obstetrics 
and Gynecology 5,12,4281--4284.

\qparr
Von Neumann (J.) 1956: Probabilistic logics and synthesis of reliable
organisms from unreliable components. In: Shannon (C.E.) and McCarthy,
(J.), editors; in Annals of Mathematical Studies, no.34, p.43-98.
Princeton University Press, Princeton.\qL
[The paper corresponds to a series of lectures given by the
author at Caltech in January 1952.] 

\qparr
Wallden (M.), Fange (D.), Lundius (E.G.), Baltekin (\"O.), Elf (J.) 2016:
The synchronization of replication and division
cycles in individual E. coli cells.
Cell 166,729-739.

\qparr
Williamson (E.M.) 1965: Incidence and family aggregation
of major congenital malformations of central nervous system.
Journal of Medical Genetics 2,161-169.

\qparr
Wright (T.P.) 1936: Factors affecting the cost of airplanes. 
Journal of Aeronautical Sciences 3,4,122-128.

\qparr
Yerushalmy (J.) 1938: Neonatal mortality by order of birth and 
age of parents.
American Journal of Epidemiology 28,2,244-270.

\qparr
Yu (Y.), Cozen (W.), Hwang (A.E.), Cockburn (M.G.), Zadnick (J),
Hamilton (A.S.), Mack (T.), Figueiredo (J.C.) 2019:
Birth anomalies in monozygotic and dizygotic twins:
results from the California twin registry.
Journal of Epidemiology 29,1,18-25.

\end{document}